\documentclass[a4paper,11pt]{article}
\usepackage{jheppub} 
\usepackage{lineno}
\usepackage{bm}
\usepackage{placeins}
\usepackage[]{graphicx}
\usepackage{amsfonts}
\usepackage{float}
\usepackage{multirow}
\usepackage{graphicx,color,amsmath,amssymb}
\usepackage{array}
\usepackage[utf8]{inputenc}
\usepackage{tikz-feynman}
\usetikzlibrary{arrows.meta}
\usetikzlibrary{decorations.markings}

\usepackage[colorlinks=true, linkcolor=blue, citecolor=blue, urlcolor=blue]{hyperref}

\usetikzlibrary{decorations.pathmorphing}
\tikzset{
  gluon/.style={decorate, decoration={coil, amplitude=2pt, segment length=4pt}},
}

\tikzset{
  fermion/.style={->, line width=0.4pt},
}

\newcommand{\eg}{\textit{e.g.}}

\newcommand{\ie}{\textit{i.e.}}
\newcommand{\dd}{\mathrm{d}}
\newcommand{\lsim}{\lesssim}
\newcommand{\gsim}{\gtrsim}

\newcommand{\Qb}{\bar{Q}}
\newcommand{\QQb}{Q\bar{Q}}

\newcommand{\QQ}{\mathcal{Q}}
\newcommand{\Tm}{T\text{-matrix}}
\newcommand{\Tms}{T\text{-matrices}}

\definecolor{mypink1}{RGB}{235, 82, 247}
\newcommand{\cm}{\textrm{c.m.}}
\newcommand{\Ecm}{E_\textrm{cm}}
\newcommand{\pcm}{p_\textrm{cm}}
\newcommand{\tcm}{\textrm{cm}}
\newcommand{\pth}{p^{\text{diss}}_{\text{th}}}

\title{\boldmath Non-perturbative quarkonium dissociation rates in strongly coupled quark-gluon plasma}

\author[a]{Biaogang Wu,}
\author[a]{Zhanduo Tang,}
\author[a]{Ralf Rapp}
\affiliation[a]{Cyclotron Institute and Department of Physics and Astronomy, Texas A$\&$M University, College Station, TX 77843-3366, USA}

\emailAdd{bgwu@tamu.edu}
\emailAdd{zhanduotang@tamu.edu}
\emailAdd{rapp@comp.tamu.edu}

\abstract{
 Heavy quarks and quarkonia are versatile probes of the transport properties of the hot QCD medium produced in ultra-relativistic heavy-ion collisions (URHICs). A robust description of heavy-flavor transport coefficients requires a microscopic approach that treats the open and hidden heavy-flavor sectors on the same footing. 
 Here, we employ the quantum many-body $T$-matrix formalism to evaluate the dissociation rates of heavy quarkonia in the quark-gluon plasma (QGP). The  basic ingredient is the heavy-light $T$-matrix, which utilizes a nonperturbative driving kernel constrained by lattice-QCD data.
 Its resummation in a ladder series provides a much enhanced interaction strength compared to previously used perturbative coupling to the quasiparticle partons in the QGP. The in-medium quarkonium properties, particularly their temperature-dependent binding energies, are obtained from selfconsistent calculations with the same interaction kernel, including interference effects  (also referred to as the imaginary part of the heavy-quark potential) as well as off-shell parton spectral functions.
 We systematically investigate the interplay of these effects and elaborate on the connections to the dipole approximation used in effective field theory.
}
\begin{document}
\maketitle

\section{Introduction}
The microscopic description of the Quark-Gluon Plasma (QGP) and its transition into hadronic matter remains a central goal in nuclear physics. Heavy quarkonia provide an excellent tool in this regard, as their vacuum spectrum reflects the fundamental QCD force, which can serve as a controlled starting point to study its manifestations in a strongly interacting medium. Thus, a systematic investigation of quarkonium production in ultra-relativistic heavy-ion collisions (URHICs), where a strongly coupled QGP (sQGP) is believed to form, provides unique insights into the interactions in the medium~\cite{Matsui:1986dk,Kluberg:2009wc,Braun-Munzinger:2009dzl,Rapp:2008tf,Mocsy:2013syh,Rapp:2017chc}. 
The dynamical nature of the expanding QGP fireball requires the deployment of transport models to track the suppression and re-emergence of the bound states as the medium cools toward freezeout.
The key role is then played by pertinent transport parameters, most notably the inelastic quarkonium reaction rates.
These rates generally suppress the quarkonium abundances in the early stages of a heavy-ion reaction, but subsequently drive them toward their (temperature-dependent) equilibrium values once bound states can be supported. Both the rates and the equilibrium limits encode the information on the microscopic properties of the quarkonia in medium. Extensive theoretical efforts are being conducted to quantify these properties and implement them into transport simulations in heavy-ion collisions, see, \eg, \cite{Andronic:2024oxz} for a recent survey and comparisons of different approaches.

In this paper, we conduct a detailed analysis of quarkonium reaction rates for both bottomonia and charmonia, which are the key transport parameter for their kinetics.
The rate may be thought of as consisting of two fundamental building blocks (although in practice, they are intertwined).
The first pertains to the in-medium properties of quarkonia binding, \ie, their binding energy, $E_B$ (typically defined as the energy gap to the open HF threshold), and internal structure (\eg, their sizes, $r$),
while the second characterizes their coupling to the light partons in the QGP.
The former is chiefly related to the in-medium $Q\bar Q$ interaction (usually treated in a potential approximation based on the large heavy-quark (HQ)  mass, $m_Q$),
while the latter is at the origin of the inelastic processes driven by the heavy-light (HL) interactions of the heavy quarks within the bound state. 
Many of the current transport approaches are based on a perturbative coupling of the quarkonium states to the medium,
or utilize temperature-dependent coefficients estimated from lattice QCD (lQCD), which are restricted to vanishing quarkonium momentum. 
Here we perform a microscopic calculation that accounts for both the internal quarkonium dynamics and the HL coupling
based on nonperturbative interactions and in a selfconsistent way. selfconsistency is particularly important in the presence of large interactions strength, which requires resummations and leads to nontrivial particle spectral functions. 
The need for nonperturbative interactions has been clearly established over the last decade in the open heavy-flavor (HF) sector~\cite{Rapp:2018qla},
and is further stipulated by the uncontrolled perturbative behavior of the HQ diffusion coefficient, ${\cal D}_s$, already at next-to-leading order~\cite{CaronHuot:2007gq}.  
Furthermore, phenomenological studies of quarkonium transport suggest that dissociation rates with perturbative couplings require a multiplicative factor of $K\simeq$~3-5 to account for experimentally observed suppression of the $\psi(2S)$~\cite{Du:2015wha} and of bottomonium states~\cite{Du:2019tjf} in the presence of a strong $Q\bar Q$ potential~\cite{Liu:2017qah}.

Our paper is organized as follows: In section~\ref{sec:hf-tmat}, we recapitulate the inputs and results of HQ interactions in the QGP within the $\Tm$ formalism (subsection~\ref{ssec:tmat}), as needed for the calculation of the quarkonium rates in this paper; this 
includes the constraints from lattice QCD (subsection~\ref{ssec:lqcd}), the HL interactions (subsection~\ref{ssec:hl-tmat}), the single-particle spectral functions (subsection~\ref{ssec:sf}), and a recently developed complex pole analysis of quarkonia, which allows us to extract their masses and dissociation widths at vanishing three-momentum in a rigorous way (subsection~\ref{ssec:spectroscopy}).
In section~\ref{sec:rates}, we present our results and analysis of the quarkonium rates by systematically elaborating and comparing various levels of approximations, including the quasifree (QF) approximation (subsection~\ref{ssec:qf}), a comparison of perturbative and nonperturbative matrix elements in on-shell kinematics (subsection~\ref{ssec:onshell}), followed by illustrating the role of in-medium parton spectral functions (subsection~\ref{ssec:offshell}) and interference effects (subsection~\ref{ssec:inter}) leading to our main results, and their comparison to results from the widely used dipole expansion (subsection~\ref{ssec:approx}).  
We discuss the implications of our findings and conclude in section~\ref{sec:concl}.

\section{Heavy-flavor $T$-matrices in the sQGP}
\label{sec:hf-tmat}
The fundamental quantities that characterize the properties of quarkonia in the QGP are their masses and decay widths, which are encoded in their spectral functions. However, in practice, the information from the spectral functions is not easily converted into a quantitative inelastic reaction rate which is suitable for phenomenological applications.
In this section, we lay out the basic components of previous calculations of quarkonium masses and widths in the thermodynamic $\Tm$ formalism that can be utilized toward this end.

In subsection~\ref{ssec:tmat}, we revisit the motivation and implementation of the $\Tm$ approach for describing quarkonium properties in the QGP.
In subsection~\ref{ssec:lqcd}, we discuss how the $\Tm$ can be constrained by using
lQCD data of HQ free energies (subsection~\ref{sssec:SCS}) and
Wilson line correlators (subsection~\ref{sssec:WLC}).
In subsection~\ref{ssec:hl-tmat}, we review the key properties of heavy-light $\Tms$, \ie, the HQ scattering amplitudes off light partons in the medium, and the single-particle spectral functions in subsection~\ref{ssec:sf}.
Finally, subsection~\ref{ssec:spectroscopy} recalls recent work on the extraction of quarkonium masses and widths via a pole analysis of their $\Tm$ in the complex energy plane, which provides essential inputs for subsequent calculations of charmonium and bottomonium dissociation rates at finite three-momentum.

\subsection{$T$-matrix formalism}
\label{ssec:tmat}
The thermodynamic $\Tm$ is a quantum many-body approach that can incorporate  a variety of nonperturbative effects that are believed to be essential for describing the sQGP. In particular, it can accommodate large interaction strengths encoded in the input potential by resumming the pertinent ladder series, incorporate off-shell effects through nontrivial parton spectral functions that emerge from large collisional widths encoded in in-medium selfenergies, and solve the pertinent one- and two-body correlation functions selfconsistently.
This enables a microscopic description of transport properties in a strongly coupled regime, while the emergence of bound states as temperature decreases can be viewed as a realization of a hadronization mechanism.     
The starting point is the 4-dimensional (4D) Bethe-Salpeter equation which can be reduced to a 3D $\Tm$ equation by taking advantage of the suppression in the energy transfer, $q_0$, in the scattering of heavy particles in a heat bath, $q_0 \sim q^2/2m_Q \ll q \sim T$  ($q$, $m_Q$ and $T$ denote the magnitude of three-momentum transfer, HQ mass, and temperature, respectively).   
Upon using a partial-wave expansion one arrives at a one-dimensional integral equation,
\begin{equation}
\begin{aligned}
T_{Q\bar{Q}}^{L,a} (E,p,p')=V_{Q\bar{Q}}^{{L,a}} (p,p') &+\frac{2}{\pi } \int_{0}^{\infty}k^{2}dk \ V_{Q\bar{Q}}^{{L,a}}(p,k) \ G_{Q\bar{Q}}^{0}(E,k) \ 
T_{Q\bar{Q}}^{L,a}(E,k,p') \ ,
\end{aligned}
\label{eq:Tm}
\end{equation}
where $p$ and $p'$ represent the magnitudes of incoming and outgoing momenta in the center-of-mass ($\cm$) frame, respectively, $a$ and $L$ the color and angular-momentum channels, and $E$ the total two-body energy. The integration variable, $k$, denotes the magnitude of the relative three-momentum in the intermediate quark-antiquark propagator,
\begin{equation}
G_{Q\bar Q}^0(E,{\bf k}) = \int d\omega_1 d\omega_2  \frac{\rho_Q(\omega_1,{\bf k}) \rho_{\bar{Q}}(\omega_2,{\bf k})} {E - \omega_1 - \omega_2 +i\epsilon}
\left[1 - n_Q(\omega_1) - n_{\bar{Q}}(\omega_2)\right]  \  , 
\label{eq:G2}
\end{equation}
which can be obtained as a convolution of two one-particle spectral functions,
\begin{equation}
\rho_{Q,i} = -\frac{1}{\pi} {\rm Im}~G_{Q,i} \ ,
\label{eq:sf1}
\end{equation}
with the one-particle propagator
\begin{equation}
G_{Q,i} = \frac{1}{\omega - \varepsilon_{Q,i}(\mathbf{k}) - \Sigma_{Q,i}(\omega,\mathbf{k})}\,,
\label{eq:propagator1}
\end{equation}
where $\varepsilon_{Q,i}(k) =\sqrt{m_{Q,i}^2 + k^2}$ denotes the on-shell energy of a heavy quark or a light parton ($i$). The single-parton selfenergies, $\Sigma_{Q,i}$, are calculated by closing the pertinent $\Tm$ (illustrated diagrammatically in figure~\ref{fig:hl-scattering}) with an in-medium light-parton propagator from the heat bath~\cite{Liu:2017qah} (see figure~\ref{fig:selfenergy}); they contain both real and imaginary parts, corresponding to an in-medium change in the dispersion relation and a collisional width, respectively.
In addition, the parton masses, $m_{Q,i}$, contain a non-dispersive, momentum independent contribution; for heavy quarks, this is computed from the Fock term of the potential, while for light partons, this contribution is utilized as a fit parameter for the QCD equation of state (associated with (gluon) condensate physics that is not included in the current $\Tm$ formalism)~\cite{Liu:2017qah}.
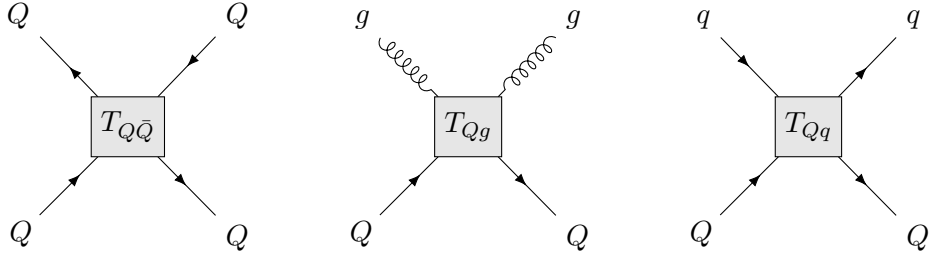
\begin{figure}[t]
   \centering
   
       \begin{tikzpicture}
       \begin{scope}[xshift=-4.5cm]
         \begin{feynman}
     \vertex (Q1) {$Q$};
     \vertex [above right=0.35cm of Q1] (QQ2);
     \vertex [above right=1.44cm of Q1] (a);
     \vertex [right=0.792cm of a] (b);
     \vertex [above=0.792cm of b] (c);
     \vertex [left=0.792cm of c] (d);
     \vertex [below right=1.08cm of b] (Q) {$Q$};
     \vertex [above right=1.08cm of c] (p2) {$\bar{Q}$};
     \vertex [above left=1.08cm of d] (e) {$\bar{Q}$};
     \draw [postaction={decorate}, decoration={
        markings,
        mark=at position 0.7 with {\arrow{Latex}}
      }] (QQ2) -- (a);
     \draw [postaction={decorate}, decoration={
        markings,
        mark=at position 0.5 with {\arrow{Latex}}
      }] (b) -- (Q);
     \draw [postaction={decorate}, decoration={
        markings,
        mark=at position 0.5 with {\arrow{Latex}}
      }] (p2) -- (c);
     \draw [postaction={decorate}, decoration={
        markings,
        mark=at position 0.5 with {\arrow{Latex}}
      }] (d) -- (e);

    \node [draw, shape=rectangle, fill=gray!20, minimum width=0.792cm, minimum height=0.792cm] at ($(a)+(0.396cm, 0.396cm)$) {{$T_{Q\bar{Q}}$}};
    \coordinate (mid) at ($(Q1)!0.2!(b)+(0,-0.072cm)$);
  \end{feynman}
  \end{scope}
   \begin{scope}[xshift=0cm]
         \begin{feynman}
     \vertex (Q1) {$Q$};
     \vertex [above right=0.35cm of Q1] (QQ2);
     \vertex [above right=1.44cm of Q1] (a);
     \vertex [right=0.792cm of a] (b);
     \vertex [above=0.792cm of b] (c);
     \vertex [left=0.792cm of c] (d);
     \vertex [below right=1.08cm of b] (Q) {$Q$};
     \vertex [above right=1.08cm of c] (p2) {$g$};
     \vertex [above left=1.08cm of d] (e) {$g$};
     \draw [postaction={decorate}, decoration={
        markings,
        mark=at position 0.7 with {\arrow{Latex}}
      }] (QQ2) -- (a);
     \draw [postaction={decorate}, decoration={
        markings,
        mark=at position 0.5 with {\arrow{Latex}}
      }] (b) -- (Q);
    \diagram* {
      (p2) -- [gluon] (c),
      (e) -- [gluon] (d),
    };
    \node [draw, shape=rectangle, fill=gray!20, minimum width=0.792cm, minimum height=0.792cm] at ($(a)+(0.396cm, 0.396cm)$) {{$T_{Qg}$}};
    \coordinate (mid) at ($(Q1)!0.2!(b)+(0,-0.072cm)$);
  \end{feynman}
    \end{scope}
  \begin{scope}[xshift=4.5cm]
         \begin{feynman}
     \vertex (Q1) {$Q$};
     \vertex [above right=0.35cm of Q1] (QQ2);
     \vertex [above right=1.44cm of Q1] (a);
     \vertex [right=0.792cm of a] (b);
     \vertex [above=0.792cm of b] (c);
     \vertex [left=0.792cm of c] (d);
     \vertex [below right=1.08cm of b] (Q) {$Q$};
     \vertex [above right=1.08cm of c] (p2) {$q$};
     \vertex [above left=1.08cm of d] (e) {$q$};
     \draw [postaction={decorate}, decoration={
        markings,
        mark=at position 0.7 with {\arrow{Latex}}
      }] (QQ2) -- (a);
     \draw [postaction={decorate}, decoration={
        markings,
        mark=at position 0.5 with {\arrow{Latex}}
      }] (b) -- (Q);
     \draw [postaction={decorate}, decoration={
        markings,
        mark=at position 0.5 with {\arrow{Latex}}
      }] (c) -- (p2);
     \draw [postaction={decorate}, decoration={
        markings,
        mark=at position 0.5 with {\arrow{Latex}}
      }] (e) -- (d);

    \node [draw, shape=rectangle, fill=gray!20, minimum width=0.792cm, minimum height=0.792cm] at ($(a)+(0.396cm, 0.396cm)$) {{$T_{Qq}$}};
    \coordinate (mid) at ($(Q1)!0.2!(b)+(0,-0.072cm)$);
  \end{feynman}
  \end{scope}
   \end{tikzpicture}
\caption{
Scattering of heavy quarks with heavy antiquarks (left), gluons (middle) and light quarks or antiquarks ($u$, $d$, $s$, right).
}
\label{fig:hl-scattering}
\end{figure}
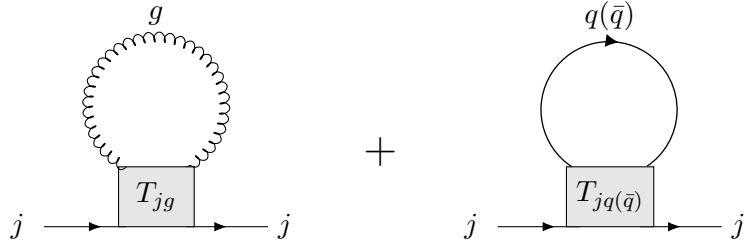
\begin{figure}[b]
  \centering
  \begin{tikzpicture}
    \begin{feynman}
      \vertex (Q1) {$j$};
      \vertex [right=0.35cm of Q1] (QQ2);
      \vertex [right=1.44cm of Q1] (a);
      \vertex [right=0.792cm of a] (b);
      \vertex [above=0.792cm of b] (c);
      \vertex [left=0.792cm of c] (d);
      \vertex [right=1.08cm of b] (Q) {$j$};
        \vertex [xshift=-0.4cm, yshift=2cm] at (c) (p2) {$g$};

      \draw[postaction={decorate}, decoration={
         markings,
         mark=at position 0.7 with {\arrow{Latex}}
       }] (QQ2) -- (a);
      \draw[postaction={decorate}, decoration={
         markings,
         mark=at position 0.5 with {\arrow{Latex}}
       }] (b) -- (Q);

      \node[draw, shape=rectangle, fill=gray!20,
            minimum width=1cm, minimum height=0.792cm]
        at ($(a)+(0.396cm,0.396cm)$) {$T_{jg}$};
    \end{feynman}

    \draw[gluon]
  (d)
  arc[start angle=244, end angle=-56, radius=0.9cm]
  -- (c);

\node at (4.8,1) {\LARGE $+$};

    \begin{scope}[xshift=6cm]
      \begin{feynman}
        \vertex (Q1) {$j$};
        \vertex [right=0.35cm of Q1] (QQ2);
        \vertex [right=1.44cm of Q1] (a);
        \vertex [right=0.792cm of a] (b);
        \vertex [above=0.792cm of b] (c);
        \vertex [left=0.9cm of c] (d);
        \vertex [right=1.08cm of b] (Q) {$j$};
        \vertex [xshift=-0.4cm, yshift=2cm] at (c) (p2) {$q(\bar{q})$};

        \draw[postaction={decorate}, decoration={
           markings,
           mark=at position 0.7 with {\arrow{Latex}}
         }] (QQ2) -- (a);
        \draw[postaction={decorate}, decoration={
           markings,
           mark=at position 0.5 with {\arrow{Latex}}
         }] (b) -- (Q);

        \node[draw, shape=rectangle, fill=gray!20,
              minimum width=0.792cm, minimum height=0.792cm]
          at ($(a)+(0.396cm,0.396cm)$) {$T_{jq(\bar{q})}$};
      \end{feynman}

    \draw[
  line width=0.5pt,
  postaction={decorate},
  decoration={
    markings,
    mark=at position 0.53 with {\arrow{Latex[width=5pt,length=5pt]}}
  }
]
(d) arc[start angle=237, end angle=-56, radius=0.9cm];

    \end{scope}

  \end{tikzpicture}
  \caption{
  Feynman diagrams for the selfenergies of partons ($j=Q,i$) in the QGP calculated by closing the in-medium $T$-matrix with a thermal gluon (left) or light quark/antiquark (right) propagator in the heat bath.
  }
  \label{fig:selfenergy}
\end{figure}
%
Furthermore, in equation (\ref{eq:G2}) $n_{Q,\Qb}$ denote the
Fermi distribution functions for the heavy (anti)quarks in thermal and chemical equilibrium (numerically they are rather negligible).

The key input to the $\Tm$ is the HQ potential, $V_{Q\bar Q}$, in equation (\ref{eq:Tm}). This quantity cannot be directly obtained from lattice QCD but rather has to be inferred from related quantities that can be calculated from the $\Tm$ and then be constrained by lQCD data. 
In practice, this is done by starting from an ansatz for the screened in-medium potential in coordinate space,
\begin{align}
\widetilde{V}(r,T) =
-\frac{4}{3} \alpha_{s} \left[\frac{e^{-m_{d} r}}{r} + m_{d}\right]
-\frac{\sigma}{m_s} \left[e^{-m_{s} r-\left(c_{b} m_{s} r\right)^{2}}-1\right] \ .
\end{align}
The coupling constant, $\alpha_s$, of the color-Coulomb interaction and the string tension, $\sigma$, of the confining force are calibrated to the vacuum HQ free energy from lQCD data~\cite{Bazavov:2018wmo}. The in-medium parameters, \ie, the screening masses, $m_{d}$ and $m_{s}$, along with an ``effective string breaking parameter'', $c_b$\footnote{$c_b$ is not obtained from a dynamical string breaking mechanism, but merely serves as a parameter to accelerate the flattening of the potential at large distances toward its asymptotic value.},
have been determined by performing selfconsistent $T$-matrix calculations for in-medium HQ free energies, Euclidean correlators, and Wilson line correlators (WLCs)~\cite{Riek:2010fk,Liu:2017qah,Tang:2023tkm} and fitting them to pertinent lQCD data.

\subsection{Lattice-QCD constraints}
\label{ssec:lqcd}
In the following, we discuss the implementation of the lQCD constraints from HQ free energies~\cite{Liu:2017qah} in subsection~\ref{sssec:SCS} 
and from WLCs as conducted in more recent work~\cite{Tang:2023tkm} in subsection~\ref{sssec:WLC}.
The latter provide a much extended dynamical reach through their dependence on Euclidean time, $\tau$, compared to the HQ free energies which are only evaluated at $\tau=1/T$. Both quantities depend on the separation, $r$, between $Q$ and $\Qb$.
  
\begin{figure}[t]
   \centering
\includegraphics[width=0.495\textwidth]{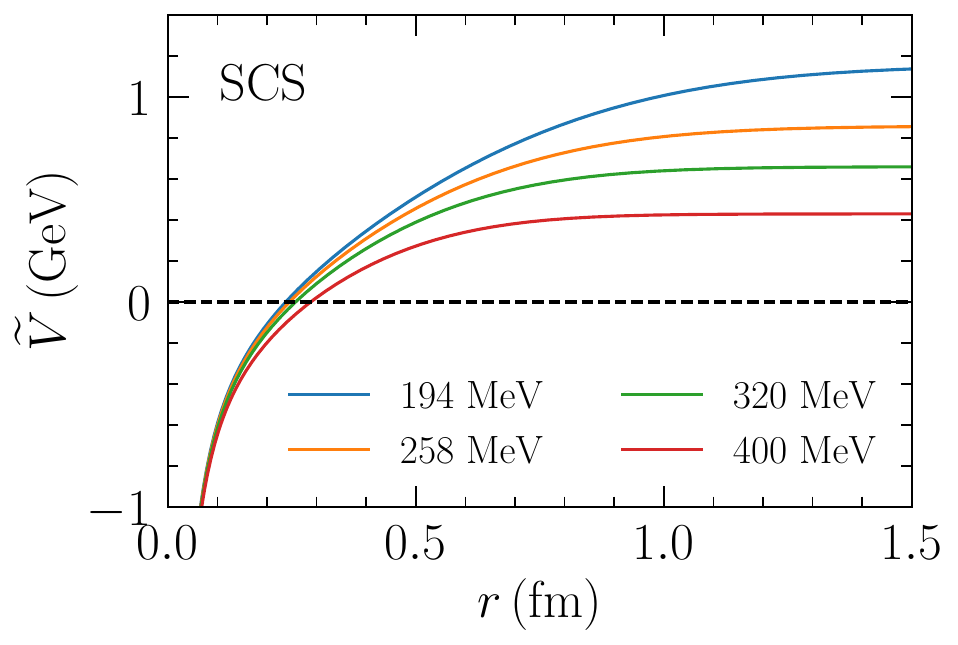}
\includegraphics[width=0.495\textwidth]{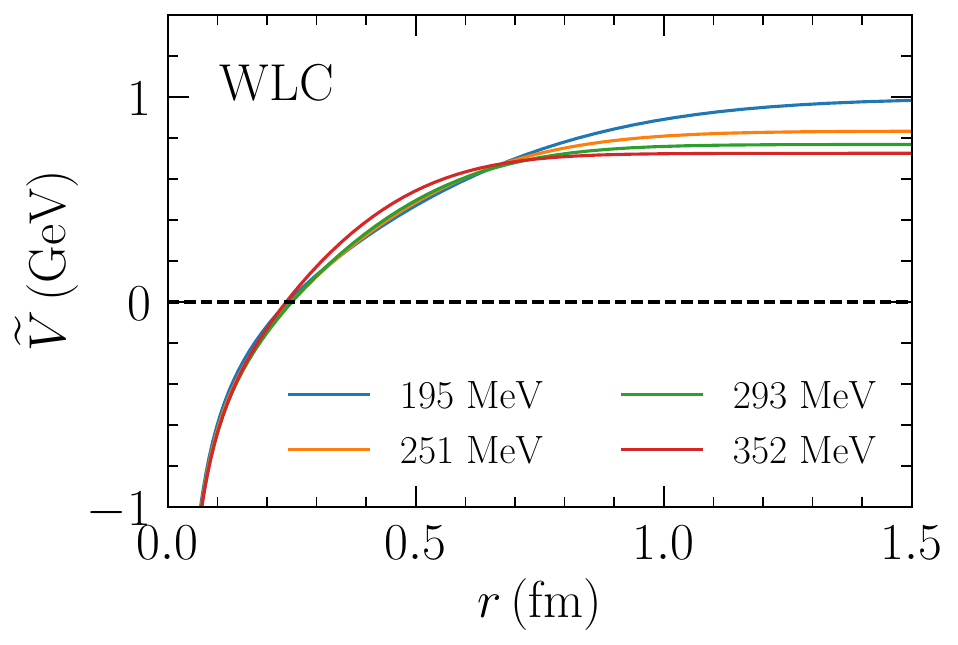}
\caption{
        The in-medium potentials $\widetilde{V}(r)$ between a heavy quark and antiquark as a function of distance $r$ at different temperatures in a strongly coupled scenario (SCS), constrained by HQ free energies (left) and in a scenario constrained by  WLCs (right).
     }
     \label{fig:potential}
  \end{figure}
\subsubsection{Strongly coupled scenario (SCS)}
\label{sssec:SCS}

High-precision lQCD data for the HQ free energy have long played an important role to better understand the $Q$-$\bar Q$ interaction in the QGP~\cite{Rapp:2008tf,Mocsy:2013syh}. While early works largely relied on approximating the HQ potential with either the free or internal energy (roughly corresponding to an adiabatic or sudden approximation, respectively)~\cite{Shuryak:2003ty}, it was subsequently realized~\cite{Liu:2015ypa} that these estimates could be systematically improved by {\em calculating} the free energy from an underlying microscopic approach and thereby constraining the potential in the spirit of a variational method.
The HQ free energy computed in lQCD is defined as the difference in the free energies of the QCD heat bath with and without a static $\QQb$ pair. 
In vacuum, this coincides with the HQ potential, $V(r)$, but at finite temperature an additional entropy term appears:
\begin{equation}
F_{\QQb}(r,T) = U_{\QQb}(r,T) - T S_{\QQb}(r,T) \ ,
\label{eq:F}
\end{equation}
which highlights the competition between minimizing the internal energy and maximizing the entropy (due to interactions with the heat bath).
To derive the corresponding expression in the $\Tm$ framework~\cite{Liu:2015ypa}, one starts 
from the definition of the HQ free energy in terms of the $\QQb$ correlation function,  
\begin{equation}
F_{\QQb}(r,T) = -T \ln \left( G^>_{\QQb} \left(-i\beta,r\right)\right) \ ,
\end{equation}
with $\beta \equiv 1/T$.
This can be elaborated as:
\begin{equation}
F_{\QQb}(r;T) = -T \ln \left[ \int\limits_{-\infty}^{\infty}- \frac{dE}{\pi} {\rm e}^{-\beta E} 
{\rm Im}\left(\frac{1}{E+i \epsilon-\widetilde{V}(r;T) -\Sigma_{\QQb}(E,r;T)} \right)\right]\ .
\label{eq:fe}
\end{equation} 
The two-body selfenergy, $\Sigma_{\QQb}(E,r;T)$, includes the interactions of the individual $Q$ and $\Qb$  with the medium (via the HQ selfenergies). In addition, its $r$ dependence encodes interference effects from three-body interactions, which turn out to be significant.
In particular, they cause an $r$-dependent suppression of the uncorrelated part of the two-body selfenergy, as derived in perturbation theory in \cite{Laine:2006ns}, and are often referred to as the imaginary part of the potential.
Diagrammatically, the underlying processes correspond to three-body diagrams, which are not easily implemented in practice.
Instead, in \cite{Liu:2017qah}, these effects were approximated in a factorized form with an interference function, $0<\phi(x)< 1$, whose form was motivated by perturbative calculations in \cite{Laine:2006ns}, $\Sigma_{\QQb}(E,r) = \Sigma_{\QQb}(E) \phi(r)$.
The expression for the free energy in (\ref{eq:fe}) has several interesting features~\cite{Liu:2015ypa}.
In the weakly coupled limit where the selfenergies vanish, the imaginary part approaches
a $\delta$-function, $-\pi\delta(E-\widetilde{V})$, which allows for direct energy integration, yielding
the result $F_{\QQb}(r;T) = \widetilde{V}(r;T)$, \ie,
for a weakly coupled system, the potential is close to the HQ free energy.
On the other hand, with large imaginary parts in the $\QQb$ selfenergy,
the spectral function is smeared out, which generally requires a stronger potential
to match the resulting HQ free energy. 
In turn, a stronger potential increases
the scattering rates in the system, requiring a selfconsistent solution to the problem.
It turns out that both types of solutions are supported in a selfconsistent determination of the potential.
However, here we focus on the ``strongly coupled scenario" (SCS),
characterized by large HQ scattering rates,
which are necessary to obtain transport parameters (such as the HQ diffusion coefficient) that are compatible with open HF phenomenology in heavy-ion collisions~\cite{Liu:2016ysz}.

The results for the HQ potential in the SCS are shown in the left panel of figure~\ref{fig:potential};
at low QGP temperatures, they feature rather little screening, with large remnants of the string interaction still present.
The latter are, in fact, pivotal in generating the large imaginary parts in the selfenergies, corresponding to collisional widths exceeding 0.5\,GeV at low momenta.
As the temperature increases, screening sets in, but the resulting widths
do not change much, as the loss in interaction strength is approximately compensated by the increase in the thermal-parton densities.
The screening mass of the string interaction is actually quite small (with a rather weak temperature dependence),  about a factor of 2-3 lower than the screening mass of the color-Coulomb potential.

\subsubsection{Wilson line correlators (WLC)}
\label{sssec:WLC}
An alternative constraint on the in-medium potential has recently been carried out by fitting the WLCs calculated from the $\Tm$~\cite{Tang:2023tkm} to lQCD data~\cite{Bala:2021fkm}. They can be written as
\begin{equation}
W\left (r,\tau,T  \right )=\int_{-\infty}^{\infty}-\frac{dE}{\pi} e^{-E \tau}{\rm Im}\left(\frac{1}{E+i \epsilon-\widetilde{V}(r;T) -\Sigma_{\QQb}(E,r;T)} \right)\ ,
\label{wlc}
\end{equation}
where the notation is identical to that used in equation (\ref{eq:fe}). The focus in \cite{Tang:2023tkm} was on the first-order cumulant of the WLCs, defined as $m_1(r, \tau, T)=-\partial_\tau \ln W(r, \tau, T)$~\cite{Bala:2021fkm}. This quantity is often interpreted as an effective mass and is widely utilized in lQCD studies; its slope with respect to $\tau$ characterizes the interacting strength between $Q\bar{Q}$ pair and the medium.
The $\Tm$ used in this study is an improved version that includes contributions from spin-dependent forces to account for the hyper/fine splitting in the vacuum spectra. This, in particular, called for a Lorentz vector component in the confining potential~\cite{Szczepaniak:1996tk}. While the impact on the spectroscopy is rather moderate, the pertinent relativistic corrections lead to a rather hard dependence of the selfenergies on the HQ three-momentum\footnote{Unless otherwise specified, we refer to a 3-momentum of a particle as relative to the thermal rest frame.} (\ie, falling off weakly with increasing three-momentum) and implying an increase in the interaction strength in the medium.
The resulting potential is comparable to the SCS at low temperatures, but is much stronger at higher temperatures, cf.~the right panel of figure~\ref{fig:potential}. 
This is mostly dictated by the large slopes in the WLCs, which imply large collisional widths requiring the screening mass of the confining potential to be essentially constant with temperature.
Consequently, the QGP remains rather strongly coupled at higher temperatures.

\begin{figure}[t]
   \centering
   \includegraphics[width=\textwidth]{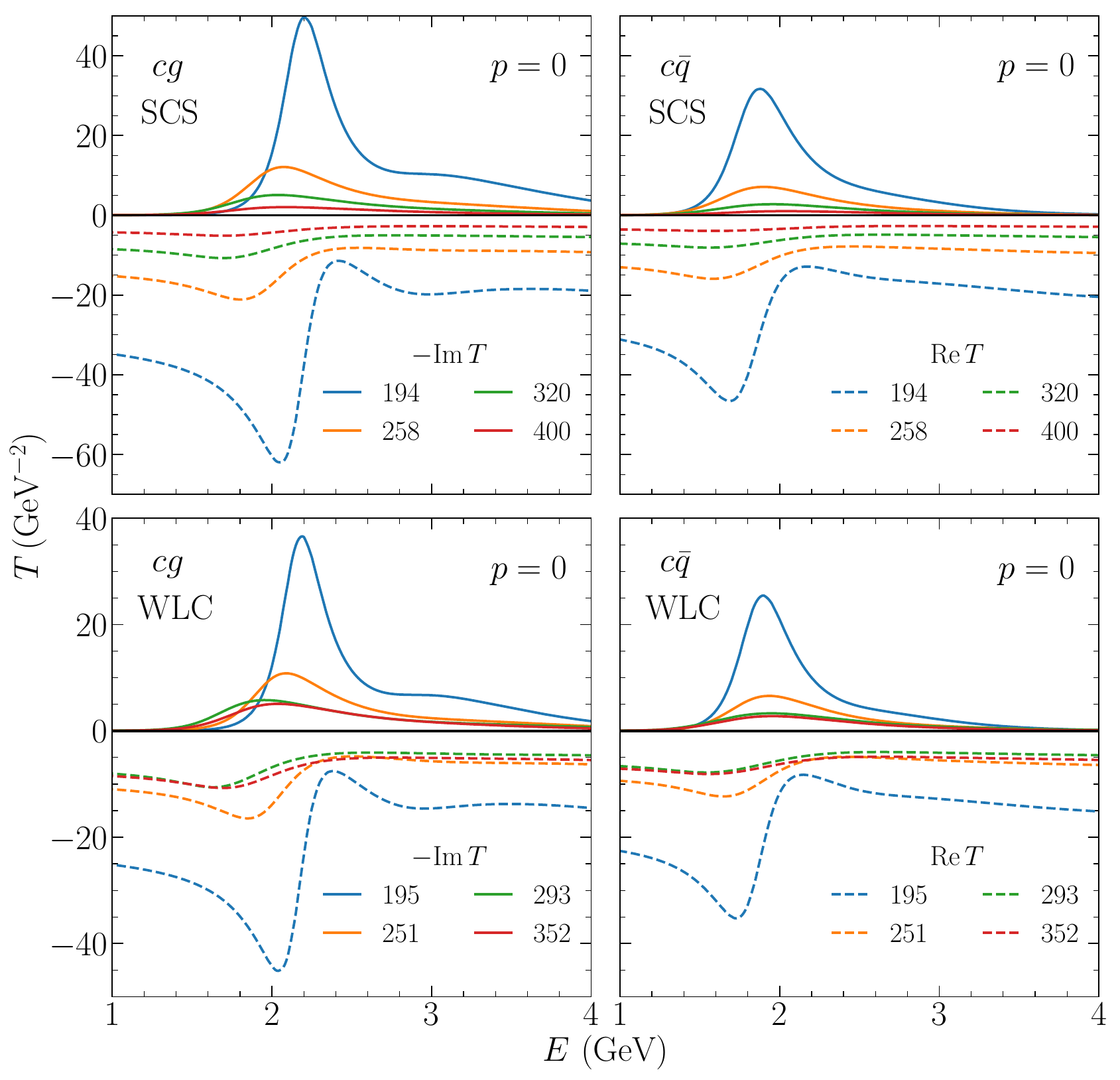}
   \caption{
The negative imaginary (solid) part and real part (dashed) of the charm-gluon $\Tm$ in the color-triplet channel, $T_{cg}$ (left), and the charm-light-quark
$\Tm$ in the color-singlet channel, $T_{c \bar q}$ (right), as functions of energy at various temperatures and zero relative momentum ($p=0$).
The upper (lower) panels correspond to the SCS (WLC) constraints.
}
     \label{fig:Tm}
  \end{figure}
%
\subsection{Heavy-light interactions}
\label{ssec:hl-tmat}
As indicated above, a central ingredient in computing inelastic dissociation rates of quarkonia in the QGP is the scattering amplitude of a heavy quark with a thermal parton, $T_{Qi}$, as depicted in the middle and right panels of figure~\ref{fig:hl-scattering}.
In the past, this was mostly evaluated using the leading order (LO) perturbative calculation from the tree-level QCD diagrams, see figure~\ref{fig:pert_diagrams}.
The $\Tm$ for heavy-light scattering, however, resums the ladder diagrams and includes the string interaction, both contributing to producing larger amplitudes.
The real and imaginary parts of $S$-wave
$cg$ scattering in the color-triplet channel and
$c\bar{q}$ scattering in the color-singlet channel (which provide the largest contribution) in the SCS and WLC scenario are displayed in figure~\ref{fig:Tm}.
The most prominent features are the large resonance structures dynamically generated at relatively low temperatures, where the resonance widths are essentially from the broad spectral functions of the re-scattering partons. 
We also see that with the WLC constraints the interaction strength decreases slower than in the SCS for temperatures above $\approx 250\,$MeV.
\begin{figure}[t]
   \centering
      \includegraphics[width=1\textwidth]{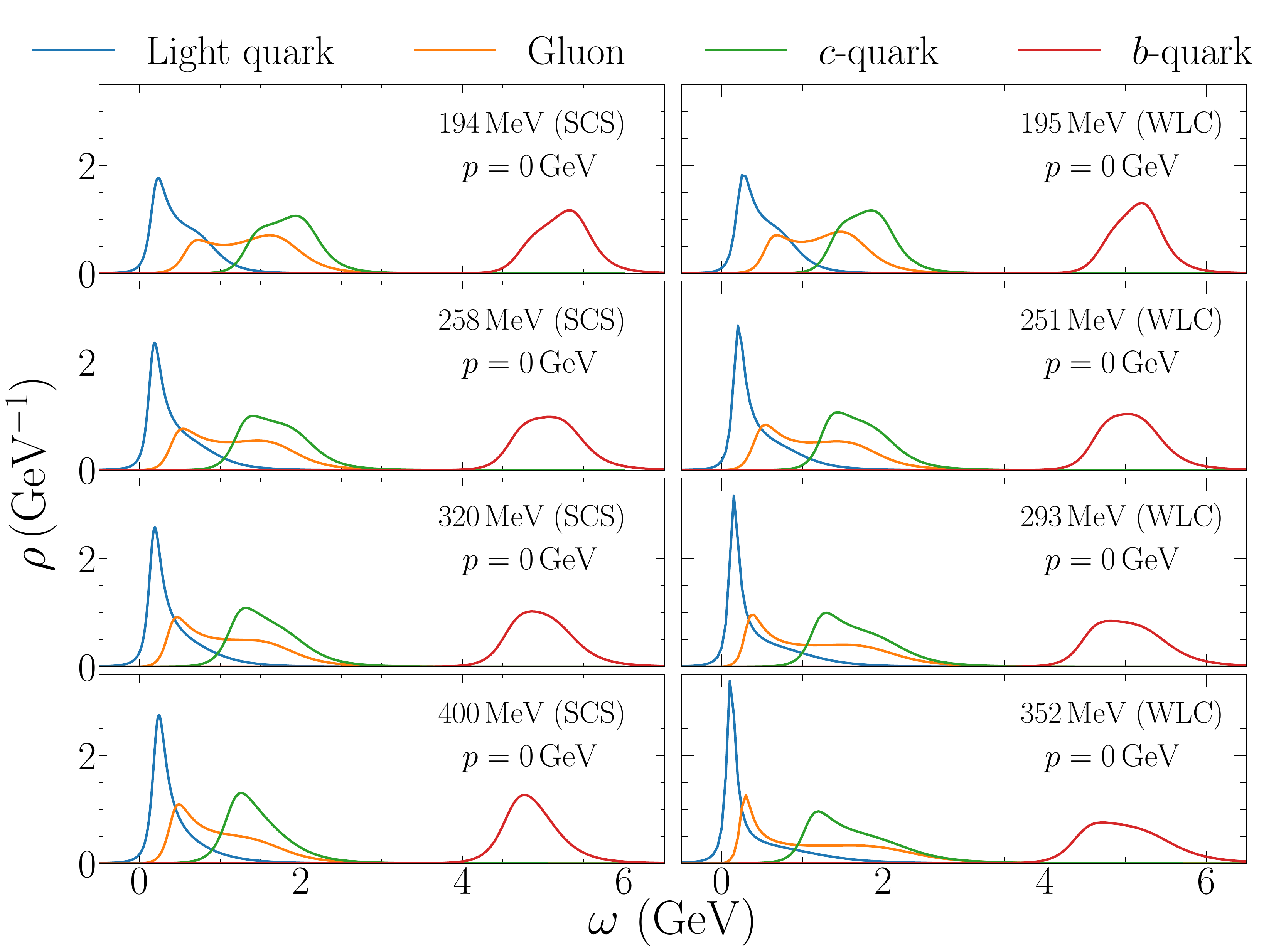}
     \caption{Spectral functions of heavy quarks and light partons at $p=0$ in the QGP for increasing temperature from top to bottom, in the SCS (left panels) and WLC scenario (right panels).}
     \label{fig:sfs_partons}
\end{figure}

\subsection{Single-particle spectral functions}
\label{ssec:sf}

\begin{figure}[t]
   \centering
      \includegraphics[width=0.99\textwidth]{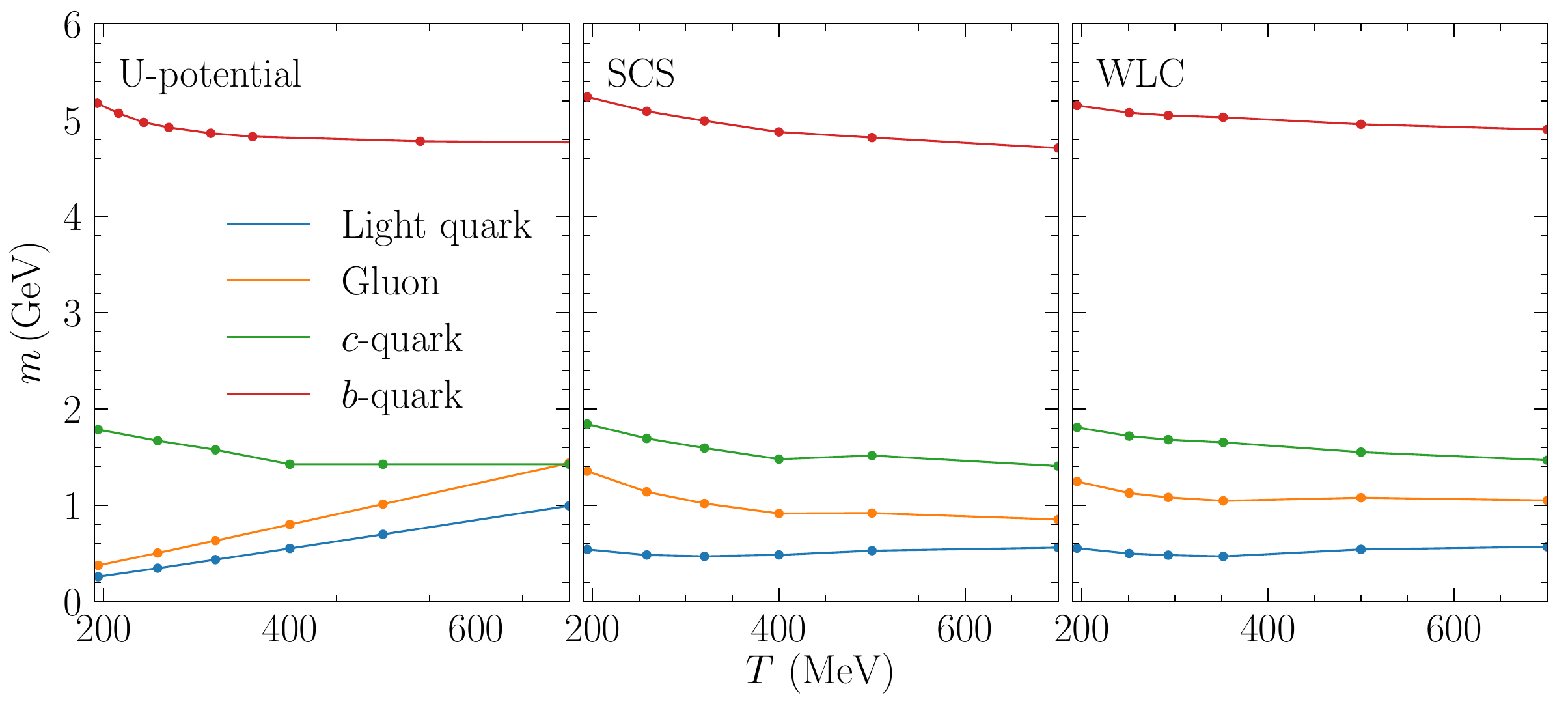}
      \caption{Masses, $m_{i,Q}$, of gluons (orange) and light (blue), charm (green) and bottom (red) quarks, as a function of temperature in the QGP for the $U$-potential (left), SCS (middle) and WLC scenario (right).}
      \label{fig:masses}
\end{figure}

Next, we turn to the single-parton spectral functions, which are selfconsistently evaluated from the HL $T$-matrices discussed in the previous section.
First, we specify the temperature-dependent HQ masses, which follow from the infinite-distance limit of the underlying potential, as $m_{Q}=\widetilde{V}(r\rightarrow\infty)/2+m^0_{Q}$, which includes a bare mass, $m^0_{Q}$ (fixed in vacuum), and a selfenergy part (``Fock term").
In the SCS, the bare masses are $m^0_{c,b}$=1.264 and 4.662\,GeV for charm and bottom, respectively~\cite{Riek:2010fk}.
In the WLC scenario, the HQ mass in vacuum is taken as $-\frac{1}{2} \int \frac{\dd^{3} \mathbf{p}}{(2 \pi)^{3}} V_{Q  \bar{Q}}^{a=1}(\mathbf{p})$ from the color-singlet potential (for $a=1$), which also reduces to $\widetilde{V}(r\rightarrow\infty)/2$ in the infinite-mass limit. The pertinent bare masses, $m^0_{c,b}$=1.359, 4.681\,GeV, are determined by fitting the full vacuum charmonium and bottomonium spectroscopy, respectively, including spin-dependent interactions~\cite{Tang:2023lcn}.
The in-medium parton masses as constrained by the lQCD equation of state, are summarized in figure~\ref{fig:masses} for the two lQCD-based scenarios.
Also shown are our earlier inputs, where the internal-energy potential ($U$) was used to obtain the quarkonium binding energies~\cite{Riek:2010fk}, along with quasiparticle parton masses, $m \sim gT$, and Born amplitudes for the HL coupling~\cite{Zhao:2010nk,Wu:2024gil}.
The analytic part of the single HQ selfenergies (other than the Fock term) is obtained from standard methods
by closing off the heavy-light forward-scattering $\Tms$ with thermal-parton propagators (recall figure~\ref{fig:selfenergy}).
Thus, the spectral properties of the light partons feed back into the selfenergies; in particular, their collisional widths play a key role in accessing the interaction strength from HL resonances, which are usually located below the nominal HL threshold (defined by the sum of the in-medium masses).

The selfconsistent results for the parton spectral functions at vanishing three-momentum for all flavors (and gluons)  are collected in figure~\ref{fig:sfs_partons} for both the SCS and WLC scenario.
The most prominent feature is a large broadening around the ``quasiparticle" peak, caused by the large imaginary parts of the parton selfenergies (here we refer to the ``quasiparticle" or nominal mass as the bare mass plus Fock term, \ie, the part without the selfenergy, as plotted in figure~\ref{fig:masses}).
For the light quarks, the width (generated by the imaginary part of the selfenergy) can exceed the nominal quasiparticle mass, which, combined with an attractive real part of the selfenergy, leads to the development of a collective mode, signaled by a low-lying peak around 0.2\,GeV, well below the nominal quasiparticle mass. Also, gluon spectral functions tend to develop this peak, especially toward higher temperature (and more so in the WLC scenario), despite their much larger quasiparticle masses compared to the light quarks. 
The charm spectral functions still exhibit a remnant of this effect in the form of a low-energy shoulder, which becomes even less significant for bottom quarks. 
One may conclude that the thermal partons are no longer good quasiparticles in the QGP, whereas charm and especially bottom quarks remain much better defined.

Another feature of the WLC scenario (not shown) is that the spectral widths do not fall off with parton three-momentum as much as in the SCS. This is a consequence of the vector component in the confining interaction, which arises due to relativistic corrections.

\subsection{Heavy-quarkonium spectroscopy}
\label{ssec:spectroscopy}
In this section, we discuss our determination of the binding energies and widths of various quarkonium states at vanishing three-momentum. These quantities are essential inputs for calculating the quarkonium dissociation rates at finite three-momentum (as needed in transport simulations), which will be discussed in section~\ref{sec:rates}.

\begin{figure}[t]
   \centering
      \includegraphics[width=0.99\textwidth]{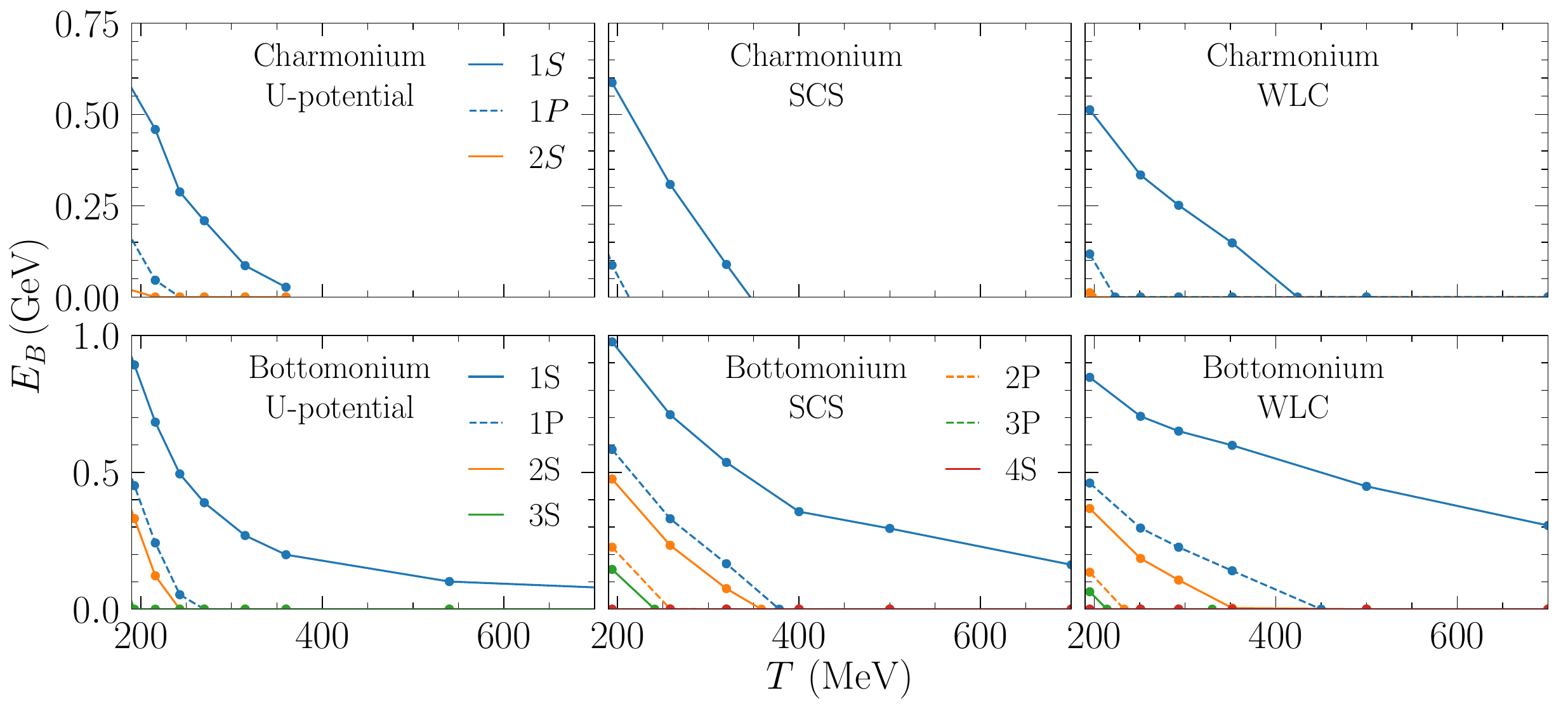}
      \caption{The binding energy, $E_B$, of charmonia (upper panels) and bottomonia (lower panels) as a function of temperature. The left panels correspond to the $U$-potential scenario, the middle panels represent the SCS, and the right panels show the WLC scenario. Solid and dashed lines represent the $S$-wave ($1S$, $2S$, $3S$, $4S$) and $P$-wave states ($1P$, $2P$, $3P$), respectively.}
    \label{fig:binding}
  \end{figure}

A conventional approach to extracting the mass and width of quarkonia involves analyzing their spectral functions. However, this method is effective only when the latter feature well-defined peaks that are amenable to fitting a localized Breit-Wigner function. As the temperature increases and a bound state begins to dissolve, its pertinent peak broadens and eventually merges into the $\QQb$ continuum, rendering mass and width extractions unreliable at best~\cite{Liu:2017qah,Tang:2024dkz}.  
To resolve this problem, we have developed a pole analysis of
the $\Tm$ in the complex energy plane~\cite{Tang:2025ypa},  
by extending the two-particle energy as $E \to z = E_R - iE_I$, with $E_I > 0$.
A pole in the $\QQb$ $T$-matrix, whose solution can be schematically written in operator form as $T(z)= V / [1-G_{2}(z)V]$, is characterized by the vanishing of both the real and imaginary parts in the denominator, and thus signals the presence of a state at $z^{\text{pole}} = E_R^{\text{pole}} -iE_I^{\text{pole}}$.
The presence of a zero in the real part critically hinges upon a sufficiently strong potential, akin to the standard bound-state solution in vacuum with a mass $M_{\cal Q} = E_R^{\text{pole}}$ (where ${\cal Q}$ denotes a quarkonium state). On the other hand, the vanishing of the imaginary part implies a width of the state given by  $\Gamma_{\cal Q} = -2E_I^{\text{pole}}$. 
However, given the large in-medium widths of the HQ spectral functions (recall figure~\ref{fig:sfs_partons}), the notion of a two-particle threshold, and thus of a binding energy, is no longer well defined. Nevertheless, as an operational definition, we will employ an effective $\QQb$ threshold, $E_{\rm thr}^{\rm eff}= 2m_Q^{\rm eff}(T)$, by evaluating an effective HQ mass, $m_Q^{\rm eff}$, at each temperature as the average over its in-medium spectral functions. 
This, in turn, will allow us to evaluate an effective binding energy in the usual way, $E_B=E_{\rm thr}^{\rm eff}-M_{\cal Q}$, where we have defined $E_B$ as positive for a state below the nominal in-medium threshold.
\begin{figure}[t]
   \includegraphics[width=0.99\textwidth]{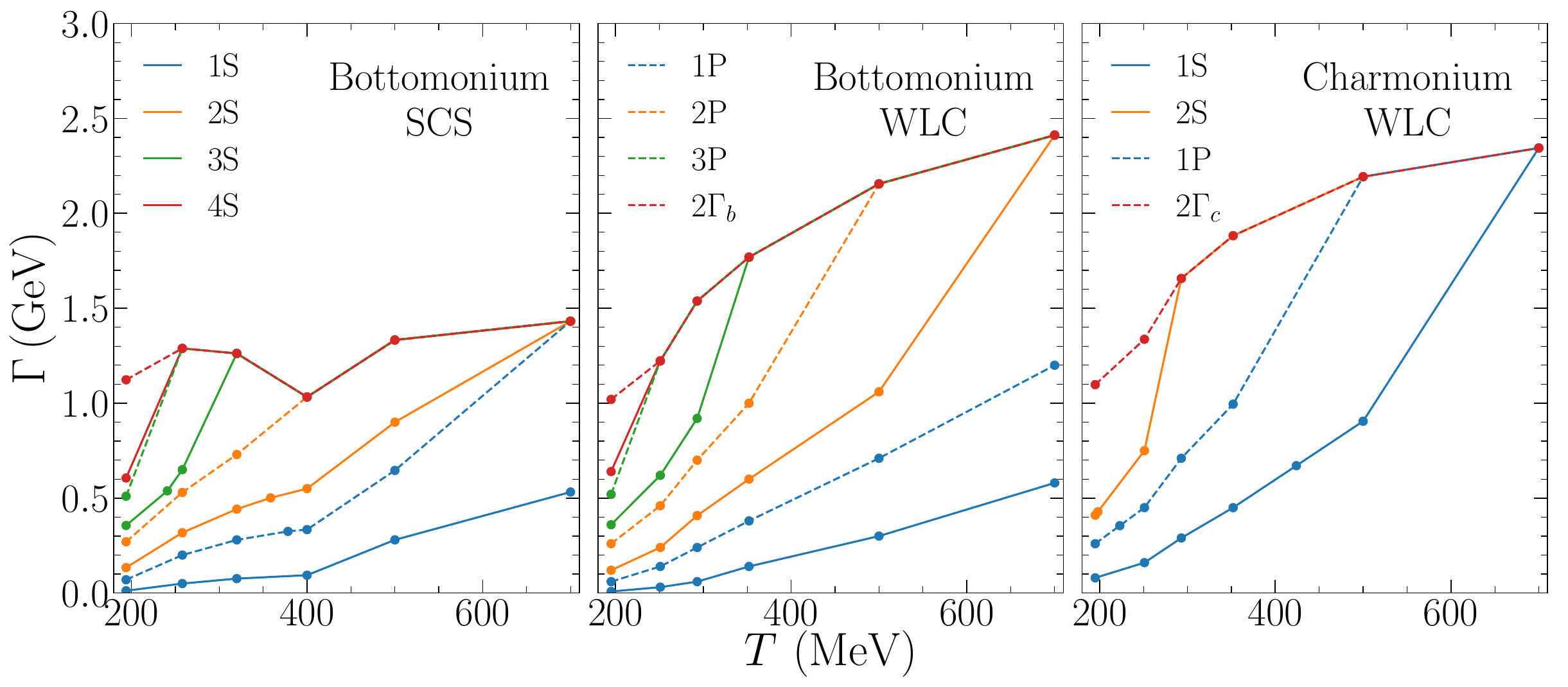}
   \caption{The thermal widths of bottomonia in the SCS (left panel) and WLC scenario (middle panel), and charmonia in the WLC scenario (right panel), as a function of temperature at vanishing quarkonium three-momentum. Solid and dashed lines denote the $S$-wave ($1S$, $2S$, $3S$, $4S$) and $P$-wave states ($1P$, $2P$, $3P$), respectively. The red dashed line indicates two times the $b$-quark width.}
    \label{fig:quarkonium_width}
\end{figure}
Figures~\ref{fig:quarkonium_width} and~\ref{fig:binding}
summarize the bottomonium  widths and binding energies for the various states as extracted from the complex-pole analysis for the two scenarios discussed above, \ie, the SCS and WLCs, as well as for charmonia for the latter.
A key feature common to both scenarios is that the melting of the bottomonium states, defined as where the pole of the $T$-matrix first disappears, occurs at temperatures well beyond the point where the nominal binding energy vanishes or becomes comparable to the width; both prescriptions have been previously used in the literature to define the melting temperature.
In the WLC scenario, the melting temperatures for the $2S$, $3S$, $4S$, $2P$, and $3P$ states are approximately 700 MeV, 350 MeV, 250 MeV, 500 MeV, and 250 MeV, respectively; the $1S$ and $1P$ states persist even at 700 MeV, which is the highest temperature considered in this study.
In contrast, in the SCS, the melting temperatures for the $2S$, $3S$, $4S$, $1P$, $2P$, and $3P$ states are approximately 700 MeV, 320 MeV, 260 MeV, 700 MeV, 400 MeV, and 260 MeV, respectively, while the $1S$ state also persists even at 700 MeV. Once a state melts, its width coincides with twice the quark width (full-width-half-maximum of the HQ spectral functions).
At low temperatures, the widths and binding energies in the SCS are comparable to those in the WLC scenario. However, at high temperatures, the binding energies and widths in the SCS are significantly smaller. This difference arises because at high temperatures, the potential in the SCS is more strongly screened than in the WLC scenario, as discussed in subsection~\ref{ssec:lqcd}.

We have also compared the results for the temperature-dependent bottomonium rates at vanishing three-momentum in the WLC scenario (taken from previous work~\cite{Tang:2025ypa}) to recent extractions from lQCD simulations with extended operators, which are limited to relatively low temperatures, $T\lsim 250$\,MeV~\cite{Ding:2025fvo}.  Within the uncertainties of the extraction method, an approximate agreement is found, but significant deviations persist for some states. As stated in \cite{Tang:2025ypa}, it remains to be understood to what extent results from correlation functions with extended operators should, in principle, agree with those from point operators (see also \cite{Andronic:2024oxz} for a more extensive comparison of calculations from various research groups).

The in-medium binding energies are compiled in figure~\ref{fig:binding}
for the two lQCD-based scenarios (middle and right panels) and compared to previous results obtained from calculations with the HQ internal energy ($U$) as the potential (left panel), while keeping the heavy-quarkonium mass constant as a function of temperature~\cite{Du:2017qkv,Wu:2024gil}.
One observes an increase in binding energies from the $U$-potential to the SCS and then to the WLC scenario, particularly toward higher temperatures, consistent with the weakest screening in the latter. This trend also aligns with quarkonium states surviving to the highest temperatures in the WLC scenario. At the lowest temperature shown ($T$=195\,MeV), all scenarios yield binding energies close to their vacuum values.

\section{Heavy-quarkonium dissociation rates}
\label{sec:rates}

We now proceed to compute the quarkonium dissociation rates to be used in transport calculations in upcoming work, employing the information we have detailed in the preceding section. We focus on dissociation processes induced by inelastic thermal-parton scattering off quarkonia. In earlier works, based on a perturbative coupling to the medium, this process was found to be dominant in practical applications, as gluo-dissociation processes are usually larger only in temperature regions where the overall magnitude of the rates is small. This is expected to be even more so when resummed scattering amplitudes are employed. 
In subsection~\ref{ssec:qf}, we introduce the quasifree approximation for the inelastic thermal-parton dissociation reactions, which simplifies the 2-to-3 process to a 2-to-2 process; in particular, we discuss two approximation schemes related to the $\cm$ motion for implementing the binding energy effect.
In subsection~\ref{ssec:onshell}, we calculate the quarkonium rates in on-shell approximation for the scattering partons, including a comparison of the $\Tm$ results with those using a perturbative amplitude.
In subsection~\ref{ssec:offshell}, we examine the effects of using off-shell spectral functions for thermal partons and the outgoing heavy quark as dictated by a consistent treatment of the quantum many-body physics.
In subsection~\ref{ssec:inter}, we implement and 
analyze interference effects that arise from the scattering of a parton off the heavy quark and antiquark inside the quarkonium.
Finally, in subsection~\ref{ssec:approx}, we scrutinize approximation schemes that have been employed in effective field theory (EFT) approaches in the dipole expansion of quarkonium structure effects and compare them with our results.

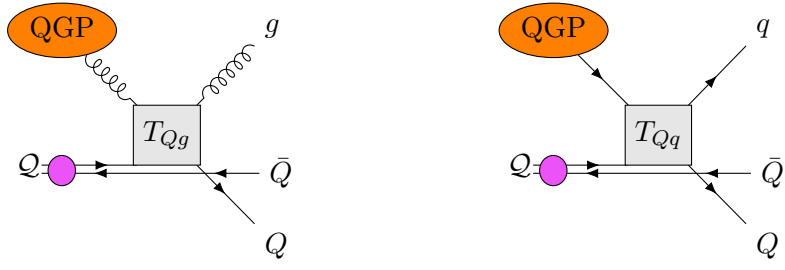
\begin{figure}[t]
   \centering
       \begin{tikzpicture}
         \begin{feynman}
     \vertex (Q1) {$\QQ$};
     \vertex [right=0.18cm of Q1] (QQ2);
     \vertex [below=0.108cm of QQ2] (Q2);
     \vertex [right=1.44cm of Q1] (a);
     \vertex [right=0.792cm of a] (b);
     \vertex [above=0.792cm of b] (c);
     \vertex [left=0.792cm of c] (d);
     \vertex [below right=1.08cm of b] (Q) {$Q$};
     \vertex [above right=1.08cm of c] (p2) {$g$};
     \vertex [right =2.88cm of Q2] (Qbar) {$\bar{Q}$};
     \vertex [blob, shape=ellipse, fill=orange, minimum size=1.44cm,minimum height=0.72cm] at ($(d)+(-0.99cm,0.99cm)$) (e) {};
     \draw [postaction={decorate}, decoration={
        markings,
        mark=at position 0.7 with {\arrow{Latex}}
      }] (QQ2) -- (a);
      \draw [postaction={decorate}, decoration={
        markings,
        mark=at position 0.21 with {\arrow{Latex}},
        mark=at position 0.76 with {\arrow{Latex}}
      }](Qbar) -- (Q2);
     \draw [postaction={decorate}, decoration={
        markings,
        mark=at position 0.5 with {\arrow{Latex}}
      }] (b) -- (Q);
    \diagram* {
      (p2) -- [gluon] (c),
      (e) -- [gluon] (d),
    };
    \node  at (e.center) {QGP};
    \node [draw, shape=rectangle, fill=gray!20, minimum width=0.792cm, minimum height=0.792cm] at ($(a)+(0.396cm, 0.396cm)$){{$T_{Qg}$}};
    \coordinate (mid) at ($(Q1)!0.2!(b)+(0,-0.072cm)$); 
    \node [draw, shape=ellipse, fill=mypink1, minimum width=0.4cm, minimum height=0.018cm, rotate=90] at (mid) {};
  \end{feynman}
  \begin{scope}[xshift=6.5cm]
         \begin{feynman}
     \vertex (Q1) {$\QQ$};
     \vertex [right=0.18cm of Q1] (QQ2);
     \vertex [below=0.108cm of QQ2] (Q2);
     \vertex [right=1.44cm of Q1] (a);
     \vertex [right=0.792cm of a] (b);
     \vertex [above=0.792cm of b] (c);
     \vertex [left=0.792cm of c] (d);
     \vertex [below right=1.08cm of b] (Q) {$Q$};
     \vertex [above right=1.08cm of c] (p2) {$q$};
     \vertex [right =2.88cm of Q2] (Qbar) {$\bar{Q}$};
     \vertex [blob, shape=ellipse, fill=orange, minimum size=1.44cm,minimum height=0.72cm] at ($(d)+(-0.99cm,0.99cm)$) (e) {};
     \draw [postaction={decorate}, decoration={
        markings,
        mark=at position 0.7 with {\arrow{Latex}}
      }] (QQ2) -- (a);
      \draw [postaction={decorate}, decoration={
        markings,
        mark=at position 0.21 with {\arrow{Latex}},
        mark=at position 0.76 with {\arrow{Latex}}
      }](Qbar) -- (Q2);
     \draw [postaction={decorate}, decoration={
        markings,
        mark=at position 0.5 with {\arrow{Latex}}
      }] (b) -- (Q);
     \draw [postaction={decorate}, decoration={
        markings,
        mark=at position 0.5 with {\arrow{Latex}}
      }] (c) -- (p2);
     \draw [postaction={decorate}, decoration={
        markings,
        mark=at position 0.5 with {\arrow{Latex}}
      }] (e) -- (d);

    \node  at (e.center) {QGP};
    \node [draw, shape=rectangle, fill=gray!20, minimum width=0.792cm, minimum height=0.792cm] at ($(a)+(0.396cm, 0.396cm)$){{$T_{Qq}$}} ;
    \coordinate (mid) at ($(Q1)!0.2!(b)+(0,-0.072cm)$); 
    \node [draw, shape=ellipse, fill=mypink1, minimum width=0.4cm, minimum height=0.018cm, rotate=90] at (mid) {};
  \end{feynman}
  \end{scope}
   \end{tikzpicture}
\caption{
Diagrammatic representation of heavy-quarkonium dissociation processes in the QGP. A heavy quarkonium, $\QQ$, is dissociated into a heavy quark, $Q$, and antiquark, $\bar{Q}$, by a thermal parton, $i$, in the QGP. The thermal parton scattering off one of the heavy quarks in the heavy quarkonium is characterized by a half-off-shell amplitude, $T_{Qi}$, that incorporates the kinematics of a finite quarkonium binding energy, $E_B$.
}
\label{fig:qq_disso}
\end{figure}

\subsection{Quasifree approximation}
\label{ssec:qf}
Initial calculations of heavy-quarkonium dissociation in hot QCD matter was conducted using the gluo-dissociation mechanism~\cite{Bhanot:1979vb}, 
\begin{equation}
   \QQ+g\rightarrow Q+\bar{Q} \ ,
\end{equation}
which, formally, is the leading order in $\alpha_s$ for the quarkonium-medium coupling (sometimes also referred to as a singlet-to-octet transition)~\cite{Brambilla:2011sg}. 
However, as the binding energy decreases (either due to screening or for excited states), gluo-dissociation rates are suppressed due to the restricted available phase space~\cite{Grandchamp:2001pf}. This suppression is further exacerbated if the gluons carry a thermal mass. As a result, 
inelastic scattering process takes over, 
\begin{equation}
   i+\QQ\to i + Q + \bar{Q} \ ,
\end{equation}
where $i=q,\bar{q}$, $g$ denotes thermal quarks, antiquarks, and gluons, respectively~\cite{Grandchamp:2001pf}.
For charmonia, explicit calculations show that gluo-dissociation is only competitive with the inelastic-scattering reactions at rather small temperatures, where both rates are small and practically rather irrelevant~\cite{Wu:2024gil}; a similar feature has been found for bottomonia as well~\cite{Du:2017qkv}. We will therefore focus on inelastic scattering in the remainder of this work.

A diagrammatic illustration of inelastic dissociation, $\mathcal{M}_{i \mathcal{Q} \rightarrow i\QQb}$, is  given in 
figure~\ref{fig:qq_disso}; the pertinent rate can be expressed as~\cite{He:2022ywp}
\begin{equation}
\begin{aligned}
   \Gamma_{\QQ}\left(\mathbf{P} ; T\right) =
   & \frac{1}{E_{\QQ}} \sum_{i}\int \dd^3 \tilde{p}_Q' \dd^3 \tilde{p}_{\bar{Q}}
   \dd^3\tilde{q} \dd^3\tilde{q}'
   \overline{\left|\mathcal{M}_{i \QQ \rightarrow i\QQb}\right|^2}
   {\left( 2\pi \right)^4}
   \delta^4\left(p_{\rm in}-p_{\rm out}\right)\\
  &\left[ 1- n_Q\left({p'}_Q^0\right) \right]\left[ 1- n_{\bar{Q}}\left({p}^0_{\bar{Q}}\right) \right] d_i n_i\left(q^0\right)\left[1 \pm n_i\left({q'}^{0}\right)\right] \ .
\end{aligned}
\label{eq:qfrate}
\end{equation}
Here, $\overline{|\mathcal{M}|^2}$ denotes the spin-color averaged squared matrix element for the $2\to3$ scattering process;
the Lorentz-invariant phase space element is defined as
\begin{equation}
   \dd^3 \tilde{p}= \frac{\dd^4 p}{\left( 2\pi \right)^3} \delta\left(p^2-m^2\right) \theta\left(p^0\right) \ ,
\end{equation}
with $m$ being the particle mass (thermal parton or heavy quark), $\mathbf{P}$ the quarkonium's three-momentum, and
\begin{equation}
    E_{\QQ}=\sqrt{M^2+\mathbf{P}^2}
\end{equation}
is the energy of the heavy quarkonium $\QQ$ of mass $M$; $d_Q$=6 denotes the spin-color degeneracy factor of the heavy quark. 
The degeneracies of thermal partons are $d_{q}=6$ for light quarks ($q$ = $u$, $d$, $s$), and $d_{g}=16$ for gluons. 
The thermal distribution functions for heavy quarks and light partons are denoted by
$n_Q$ and $n_i$, respectively,
while $q$ ($q'$) is the four-momentum of the initial (final) light parton, and $p_Q'$ and ${p}_{\Qb}$ are the four-momenta of the outgoing heavy quark and heavy antiquark.
The $\delta$-function enforces energy-momentum conservation of the $2\to3$ process, \ie, 
\begin{equation}
   \delta^4\left(p_{\rm in}-p_{\rm out}\right)=\delta^4\left(P+q-p_Q'-q'-p_{\bar{Q}}\right) \ .
\end{equation}

\begin{table}[t]
\centering
\begin{tabular}{|c|c|c|}
\hline
 & \textbf{Rest frame} & \textbf{HL $\cm$ frame}
 \\ \hline
\textbf{Relative motion} & 
$\begin{array}{c}
   \frac{\mathbf{p}_Q}{\varepsilon_Q} = \frac{\mathbf{p}_{\Qb}}{\varepsilon_{\Qb}} = \frac{\mathbf{P}}{E_{\QQ}}\\
\end{array}$
&
$\begin{array}{c}
   \mathbf{p}_Q = \mathbf{p}_{\Qb} = \frac{\mathbf{P}}{2}\\
\end{array}$
\\ \hline
\parbox{3.5cm}{\centering \textbf{Incoming heavy quark ($Q$)}}&
$\begin{array}{c}
\text{Off-shell}\\
   \tilde{m}_Q=M-m_{\Qb}
\end{array}$
&
$\begin{array}{c}
\text{On-shell}\\
\tilde{m}_Q=m_Q
\end{array}$
\\ \hline
\parbox{4cm}{\centering \textbf{Kinematics for $2\to2$ approximation}} &
$\begin{array}{c}
   \mathbf{p}_Q=\frac{\tilde{m}_Q}{M}\mathbf{P}\\
   \varepsilon_Q=\frac{\tilde{m}_Q}{M}E_{\QQ}\\
s=\left(\varepsilon_Q+\varepsilon_i\right)^2-\left(\mathbf{p}_Q+\mathbf{q}\right)^2\\
      \Ecm=\sqrt{s}\\
\end{array}$
&
$\begin{array}{c}
   \mathbf{p}_Q = \frac{1}{2}\mathbf{P}\\
\varepsilon_Q=\sqrt{m_Q^2+\mathbf{p}_Q^2}\\
s=\left(\varepsilon_Q+\varepsilon_i\right)^2-\left(\mathbf{p}_Q+\mathbf{q}\right)^2\\
   \Ecm=\sqrt{s}+M-m_Q-m_{\Qb}\\
\end{array}$
\\ \hline
\end{tabular}
\caption{Comparison of the two quasifree approximations employed in this work. The columns correspond to different reference frames in which the binding energy is considered: the quarkonium rest frame and the HL $\cm$ frame. The rows characterize the following items (from top to bottom): 
(1) the HQ and quarkonium momenta;
(2) the treatment of binding energy at the HQ level;
(3) the heavy-light $\Tm$ inputs, including the incoming HQ momentum ($\mathbf{p}_Q$) and energy ($\varepsilon_Q$), the $\cm$ energy ($\Ecm$) and implementation of binding in the heavy-light particle system.
}
\label{tab:qf}
\end{table}

The large HQ mass, relative to the quarkonium binding energy, allows for a simplification of the $2\to 3$ process into an inelastic $2\to2$ scattering within the so-called quasifree approximation~\cite{Grandchamp:2002wp}. 
Following the procedure utilized in previous work~\cite{Grandchamp:2001pf,Zhao:2010nk,Du:2017qkv}, a thermal parton scatters off one of the heavy quarks, which is taken off-shell to incorporate the binding energy in its mass,
$\tilde{m}_Q = m_Q-E_B$, 
while the spectator quark remains on-shell with its  mass. At this level, internal-structure effects of the quarkonium beyond the binding energy are neglected, but we will return to them further below.
Momentum conservation is maintained through the $\delta$-function in
\begin{equation}
\begin{aligned}
\begin{array}{l}
\overline{\left|\mathcal{M}_{i \QQ \rightarrow i \QQb}
\left(q, P, q', p_Q', {p}_{\bar{Q}}\right)\right|^2} 
\\[8pt] \hspace{100pt}
= 2 \overline{\left|\mathcal{M}_{i Q\rightarrow i Q}
\left(q, p_Q, q', p_Q'\right)\right|^2} 
2 \varepsilon_{\bar{Q}} (2\pi)^3 
\delta^{(3)}\left(\mathbf{p}_{\Qb}+\mathbf{p}_Q- \mathbf{P}\right) \ ,
\end{array}
\end{aligned}
\label{eq:qf_amplitude}
\end{equation}
where $p_Q$ is the momentum of the incoming heavy quark.
This, in turn, enables one to determine the three-momentum of the spectator heavy quark.

As an alternative approximation in this work, we utilize a definition that better aligns with the energy and momentum variables used in the heavy-light scattering amplitude, $T(\Ecm;~\pcm,~\pcm')$ in the $\cm$ frame of the $\Tm$~\cite{Liu:2017qah} (which will be discussed in more detail in the following section). Here, the momentum of the heavy quark is taken as half of the momentum of the heavy quarkonium,
$\mathbf{p}_Q=\frac{\mathbf{P}}{2}$,   
while the effect of the binding energy is 
encoded in the energy in the $\cm$ frame, defined as 
\begin{equation}
       \Ecm=\sqrt{s}- E_B \ ,
       \label{eq:Ecm}
\end{equation}
where 
\begin{equation}
s=\left(\varepsilon_Q+\varepsilon_i\right)^2-\left(\mathbf{p}_Q+\mathbf{q}\right)^2
\end{equation}
is the invariant mass squared of the heavy-light system before the collision. 

The kinematics for both approaches are summarized in table~\ref{tab:qf}.
They yield the same rates in the limit of vanishing binding energy and also for $\mathbf{P}=0$ (the quarkonium rest frame). Consequently, in our numerical results of the dissociation rates reported below, the largest deviation occurs at large binding. For the $\Upsilon(1S)$ at a temperature of $T$=195\,MeV, this deviation amounts to up to $\sim$10\%; however, we note that under these conditions the magnitude of the rate is approximately 10\,MeV (or even lower at smaller temperatures) and is therefore irrelevant in the context of $\Upsilon$ transport in heavy-ion collisions.

\subsection{Perturbative vs. nonperturbative coupling with on-shell kinematics}
\label{ssec:onshell}
In this section, we connect with our previous results where a perturbative amplitude was used to describe the quarkonium-medium coupling, incorporating on-shell kinematics for the partons.
Accordingly, we adopt the kinematics of the first approach in table~\ref{tab:qf}.
To facilitate the evaluation of the invariant matrix element in equation (\ref{eq:qfrate}), we perform the calcuation in the $\cm$ frame of the scattering process. 

In the $\cm$ frame of a two-particle system, the momenta of the two particls, $\mathbf{p}_\tcm$ and $\mathbf{q}_\tcm$, along with the corresponding energies, $p^0_\tcm$ and $q^0_\tcm$, satisfy the relations
\begin{equation}
   \mathbf{p}_{\tcm}+\mathbf{q}_{\tcm}=0, \quad p^0_{\tcm}+q^0_{\tcm}=\Ecm \ .
   \end{equation}
By solving these equations together with the on-shell conditions for the heavy quark and the thermal parton, we determine the ampitudes of the incoming and outgoing HQ momenta ($p_{\tcm}$, $p_{\tcm}'$), and their respective energies ($\varepsilon_{Q,\tcm}$, $\varepsilon_{Q,\tcm}'$):
   \begin{equation}
   \begin{aligned}
& p_{\tcm}=\frac{\sqrt{\left(\Ecm^2-\left(\tilde{m}_Q^{2}+m_i^2\right)\right)^2-4 \tilde{m}_Q^{2} m_i^2}}{2 \Ecm} \ ,\\
& p_{\tcm}'=\frac{\sqrt{\left(\Ecm^2-\left(m_Q^{2}+m_i^2\right)\right)^2-4 m_Q^2 m_i^2}}{2 \Ecm} \ ,\\
\varepsilon_{Q,\tcm}&=\frac{\Ecm^2+\tilde{m}_Q^{2}-m_i^2}{2 \Ecm} \ ,
\quad \varepsilon_{Q,\tcm}'=\frac{\Ecm^2+m_Q^2-m_i^2}{2 \Ecm} \ .
\end{aligned}
\label{eq:pcm}
\end{equation}
Note that the incoming HQ mass, $\tilde{m}_Q<m_Q$, includes the effect of the quarkonium binding. As a result, the outgoing $\cm$ momentum decreases ($\pcm'<\pcm$) to conserve energy, thereby reducing the available phase space.

\begin{figure}[t]
\centering
\begin{tikzpicture}[baseline=(current bounding box.south)]
  \begin{feynman}
     \vertex (a);
     \vertex [right=1.44cm of a] (b);
     \vertex [above left=0.8cm and 0.8cm of a] (g1) {\(g\)};
     \vertex[below left=0.8cm and 0.8cm of a] (Q1) {\(Q\)};
     \vertex[above right=0.8cm and 0.8cm of b] (g2) {\(g\)};
     \vertex[below right=0.8cm and 0.8cm of b] (Q2) {\(Q\)};
     \diagram* {
      (g1) -- [gluon] (a),
      (Q1) -- [fermion] (a),
      (a) -- [fermion] (b),
      (b) -- [gluon] (g2),
      (b) -- [fermion] (Q2),
    };
  \end{feynman}
\end{tikzpicture}
\begin{tikzpicture}[baseline=(current bounding box.south)]
  \begin{feynman}
     \vertex (a);
     \vertex [right=1.44cm of a] (b);
     \vertex [above left=0.8cm and 0.8cm of a] (i1) {\(g\)};
     \vertex[below left=0.8cm and 0.8cm of a] (i2) {\(Q\)};
     \vertex[above right=0.8cm and 0.8cm of b] (f1) {\(g\)};
     \vertex[below right=0.8cm and 0.8cm of b] (f2) {\(Q\)};
    \diagram* {
       (a) -- [fermion] (b),
      (i1) -- [gluon] (b) -- [fermion] (f2),
      (i2) -- [fermion] (a) -- [gluon] (f1),
    };
  \end{feynman}
\end{tikzpicture}
\begin{tikzpicture}[baseline=(current bounding box.south)]
  \begin{feynman}[vertical=a to b]
    \vertex (a);
    \vertex[below=1cm of a] (b);
    \vertex[above left=0.8cm and 0.8cm of a] (i1) {\(g\)};
    \vertex[above right=0.8cm and 0.8cm of a] (i2) {\(g\)};
    \vertex[below left=0.8cm and 0.8cm of b] (f1) {\(Q\)};
    \vertex[below right=0.8cm and 0.8cm of b] (f2) {\(Q\)};
    \diagram* {
      (i1) -- [gluon] (a) -- [gluon] (i2),
      (a) -- [gluon] (b),
      (b) -- [fermion] (f2),
      (f1) -- [fermion] (b),
    };
  \end{feynman}
\end{tikzpicture}
\begin{tikzpicture}[baseline=(current bounding box.south)]
  \begin{feynman}[vertical=a to b]
    \vertex (a);
    \vertex[below=1cm of a] (b);
    \vertex[above left=0.8cm and 0.8cm of a] (i1) {\(q\)};
    \vertex[above right=0.8cm and 0.8cm of a] (i2) {\(q\)};
    \vertex[below left=0.8cm and 0.8cm of b] (f1) {\(Q\)};
    \vertex[below right=0.8cm and 0.8cm of b] (f2) {\(Q\)};
    \diagram* {
      (i1) -- [fermion] (a) -- [fermion] (i2),
      (a) -- [gluon] (b),
      (b) -- [fermion] (f2),
      (f1) -- [fermion] (b),
    };
  \end{feynman}
\end{tikzpicture}

\caption{Tree-level Feynman diagrams for HQ-gluon $t$-, $u$- and $s$-channels (first three diagrams) and heavy-light quark $t$-channel (rightmost diagram) interactions.}
\label{fig:pert_diagrams}
\end{figure}
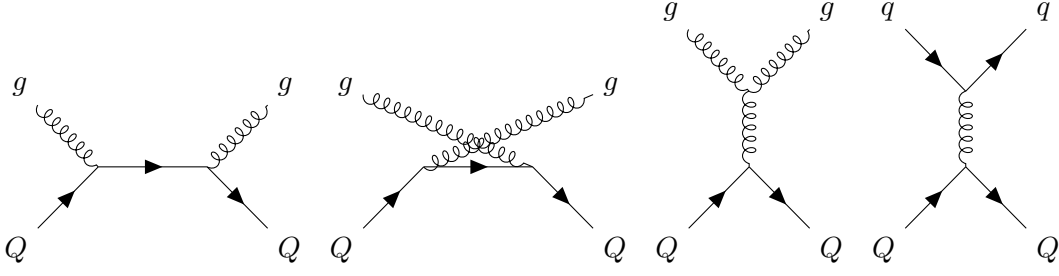

After integrating out the momenta of the spectator quark ($\mathbf{p}_{\Qb}$), the outgoing  heavy quark ($\mathbf{p}_Q'$) and the light-parton ($\mathbf{q}'$) using the energy-momentum conserving $\delta$-function, 
the quarkonium dissociation rate, equation (\ref{eq:qfrate}), simplifies to
\begin{equation}
\begin{aligned}
   \Gamma_{\QQ}\left(\mathbf{P} ; T\right)=&\frac{2}{\left( 2\pi \right)^5 2\varepsilon_Q} \sum_{i} \int \frac{q^2 \dd q \dd \Omega_{\mathbf{q}}}{2 \varepsilon_{i}}\frac{p_{\tcm}'\dd \Omega_{\tcm}}{4 \Ecm}
   \overline{\left|\mathcal{M}_{i {Q} \rightarrow i Q}\right|^2}\left[\Ecm, p_{\tcm},p_{\tcm}', \cos\theta_{\tcm} \right]\\
   &\left[ 1- n_Q\left(\varepsilon_Q'\right) \right] \left[1- n_{\bar{Q}}\left(\varepsilon_{\bar{Q}}\right) \right]d_i n_i\left(\varepsilon_i\right)
   \left[1 \pm f_p\left(\varepsilon_Q+\varepsilon_i-\varepsilon_Q'\right)\right] \ .
\end{aligned}
\label{eq:onshell}
\end{equation}
The quarkonium momentum, $\mathbf{P}$, is  related to the HQ momentum $\mathbf{p}_Q$ as specified in table~\ref{tab:qf}.
The variables $\Omega_{\mathbf{q}}$ and $\Omega_{\tcm}$ denote the solid angle of the thermal parton relative to $\mathbf{p}_Q$
and the angle between the incoming and outgoing heavy quarks in the $\cm$ frame.
The on-shell energies of the incoming heavy quark, heavy antiquark and light parton are given by
\begin{equation}
\begin{aligned}
\varepsilon_{Q}=\sqrt{\tilde{m}_Q^2+\mathbf{p}_Q^2} \ ,\qquad
\varepsilon_{\Qb}=\sqrt{m_{\Qb}^2+\mathbf{p}_{\Qb}^2} \ ,\qquad
\varepsilon_{i}=\sqrt{m_i^2+\mathbf{q}^2} \ ,
\end{aligned}
\label{eq:kinetic}
\end{equation}
where $\tilde{m}_Q$, $m_{\Qb}$ and $m_i$ are their respective masses. The energy of the outgoing heavy quark in the laboratory frame,
$\varepsilon_Q'$, is given by a Lorentz transformation,
\begin{equation}
\begin{aligned}
\varepsilon_Q' =\ & \gamma_{\tcm}\left[\varepsilon_{Q,\tcm}'+
\vphantom{\varepsilon_{\tcm}'} \left(\cos \theta_{\tcm} \mathbf{v}_{\tcm} \cdot \mathbf{p}_{\tcm}+p_{\tcm} N \sin \theta_{\tcm} \sin \phi_{\tcm}\right)\right] ,
\end{aligned}
\label{eq:E_Qf}
\end{equation}
where $\mathbf{p}_{\tcm}$ is the vector of the $\cm$ momentum and can be also derived from the Lorentz transformation,
\begin{equation}
\begin{aligned}
&\mathbf{p}_{\text{cm}} = \mathbf{p}_Q + \left[ (\gamma_{\text{cm}} - 1) \frac{ \mathbf{v}_{\text{cm}} \cdot \mathbf{p}_Q }{v_{\text{cm}}^2} - \gamma_{\text{cm}} \varepsilon_Q \right] \mathbf{v}_{\text{cm}} \,.
\end{aligned}
\end{equation}
The $\cm$ velocity, $\mathbf{v}_{\tcm}$, and corresponding Lorentz factor $\gamma_{\tcm}$ are defined as
\begin{equation}
\mathbf{v}_{\tcm}=\frac{\mathbf{p}_Q+\mathbf{q}}{\varepsilon_{Q}+\varepsilon_{i}} \ , \quad \gamma_{\tcm}=\frac{\varepsilon_{Q}+\varepsilon_{i}}{\Ecm} \ .
\end{equation}
The azimuthal angle of the incoming heavy quark and antiquark in the $\cm$ frame is defined by $\phi_{\tcm}$, and the magnitude squared of the transverse component of the outgoing HQ momentum is given by
\begin{equation}
   N^2=v_{\tcm}^2-\frac{\left(\mathbf{p}_{\tcm} \cdot \mathbf{v}_{\tcm }\right)^2}{ \mathbf{p}_{\tcm}^2} \ .
\end{equation}

\begin{figure}[t]
   \centering
    \includegraphics[width=0.445\textwidth]{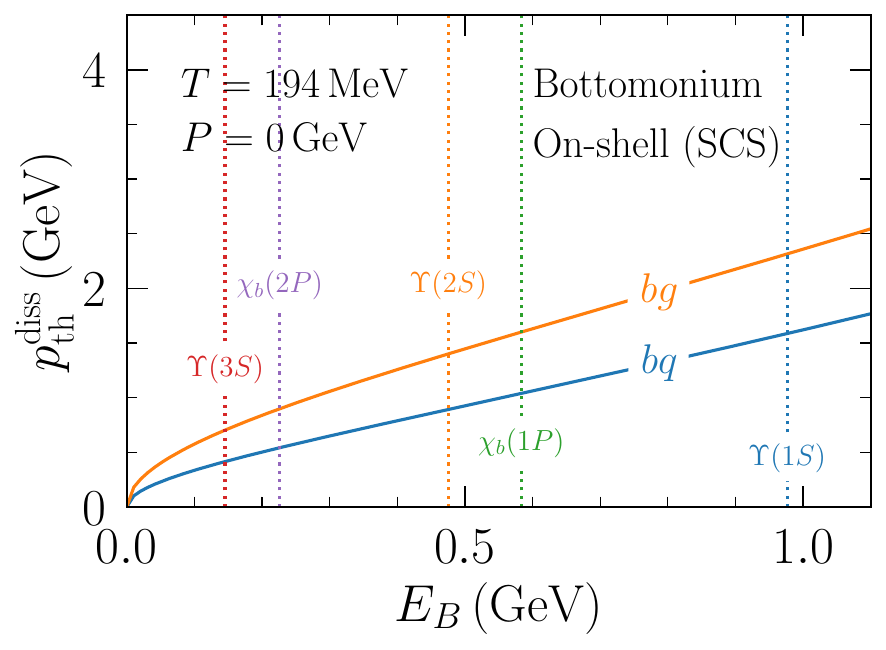}
    \includegraphics[width=0.495\textwidth]{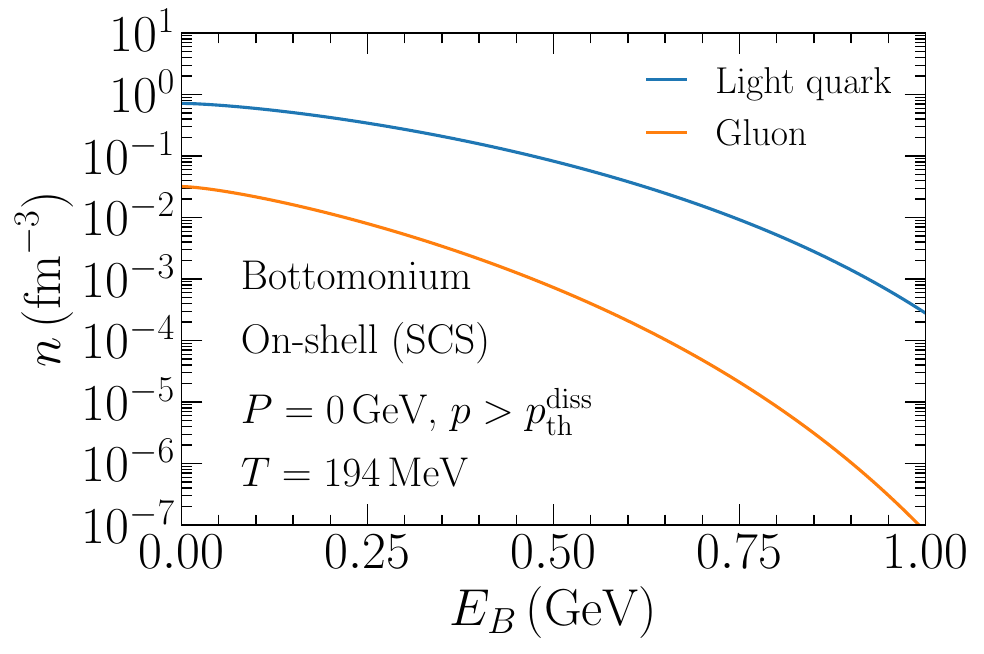}
     \caption{
Left panel: the dissociation momentum thresholds ($\pth$) of a light quark (blue) and gluon (orange) required for different bottomonia as a function of their binding energy in the QGP at $T = 194\,$MeV and $P$=0. The binding energies of the different states are taken from the SCS and are indicated by vertical dotted lines.
Right panel: number density of light quarks (blue) and gluons (orange) with  3-momenta above the threshold momentum, $p>\pth (E_B)$.
  }
     \label{fig:thresholds}
\end{figure}

Let us first quantify the effect of the binding energy. Toward this end, we calculate the threshold momentum, $\pth$, minimally required for an incoming thermal parton to dissociate the bound state, based on energy-momentum conservation:
\begin{equation}
   \left( m_Q-E_B+\varepsilon_{i,\text{th}} \right)^2-{\pth}^2=\left(m_Q+m_i\right)^2 \ ,
   \label{eq:sth}
\end{equation}
where $\varepsilon_{i,\text{th}}= \sqrt{m_i^2+{\pth}^2}$ denotes the energy of the thermal parton at threshold. 
Solving (\ref{eq:sth}) for $\pth$, we obtain
\begin{equation}
\pth=\frac{\sqrt{E_B\left(E_B+2m_i\right)\left(2m_Q-E_B\right)\left(2m_Q+2m_i-E_B\right)}}{2 \left(m_Q-E_B\right)} \ .
      \label{eq:pth}
\end{equation}
Note that charmonia generally have a higher dissociation threshold than bottomonia with the same binding energy, since {for the same $\cm$ energy $\Ecm$}, the smaller charm-quark mass results in a larger $\cm$ momentum {(see (\ref{eq:pcm}))}, thereby necessitating a faster light parton to break the bound state.
On the other hand, for the same momentum, the larger energy of a gluon relative to a light quark in the $\cm$ frame implies that gluons are less effective at breaking up quarkonia.
The left panel of figure~\ref{fig:thresholds} illustrates the dependence of the threshold momentum, required for an incoming light quark or gluon to dissociate a bottomonium at rest in a QGP medium, as a function of its binding at $T=194\,$MeV.
At this temperature, within the SCS, a light quark must have a momentum of at least $\pth\approx 1.5(1.2)\,$GeV to dissociate the $\Upsilon(1S)$ ($J/\psi$), while a gluon requires $\pth\approx 2.3\,$GeV ($\approx2.2\,$GeV for $J/\psi$). 
The suppression of the dissociation rate with increasing $E_B$ is primarily driven by the decreasing number density of thermal partons above $\pth$, given by  
\begin{equation}
n_i\left(E_B;T\right)=d_i\int_{\pth \left(E_B\right)}^{\infty}\frac{\dd^3 \mathbf{q}}{\left(2\pi\right)^3}e^{-\frac{\varepsilon_i\left(\mathbf{q}\right)}{T}} \ .
\end{equation}
As illustrated in the right panel of figure~\ref{fig:thresholds}, the number density of light partons with sufficient energy to dissociate a quarkonium  is markedly suppressed as $E_B$ increases.
At $E_B=1\,$GeV, the density of light quarks above threshold is reduced by more than three orders of magnitude, while the density of gluons decreases by over five orders of magnitude relative to the $E_B=0$ case.
\begin{figure}[t]
   \centering
   \includegraphics[width=0.495\textwidth]{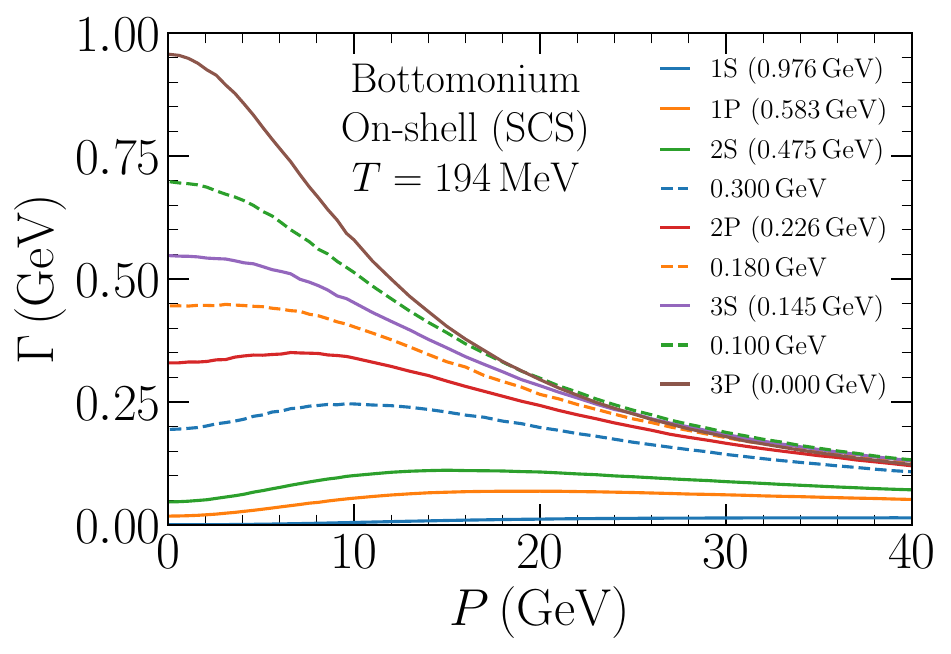}
    \includegraphics[width=0.495\textwidth]{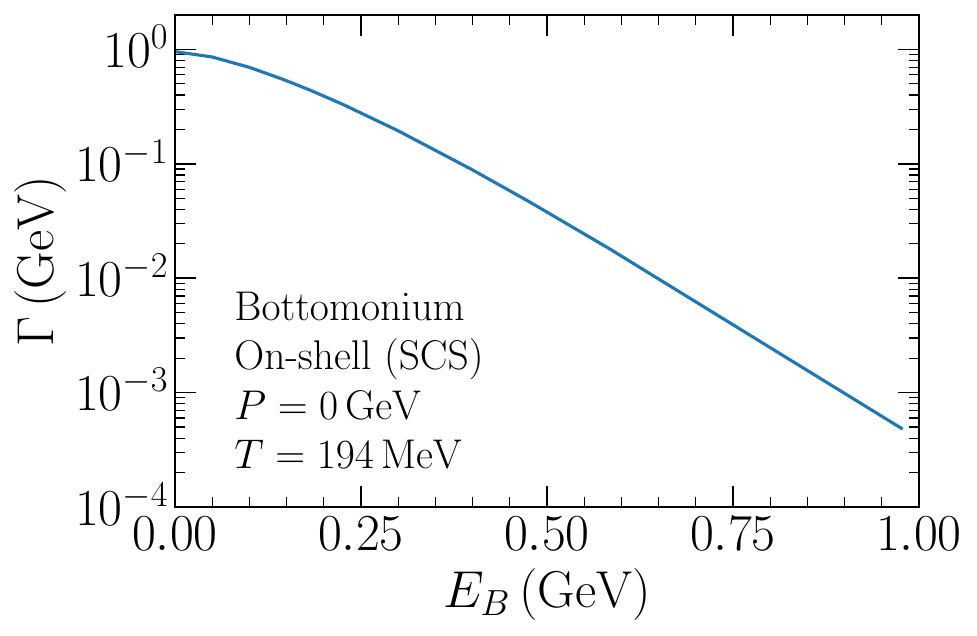}
   \caption{Left panel: dissociation rates of different bottomonium states in the SCS using on-shell partons (solid lines) and reference rates for additional binding-energy values (dashed lines) as a function of momentum $P$ at $T=194\,$MeV. Right: $\Gamma(P=0)$ as a function of $E_B$ in the on-shell SCS at $T=194$\,MeV. }
    \label{fig:b_SCS_rates_EB}
  \end{figure}
The resulting bottomonium dissociation rates as a function of 3-momemtum are compiled in figure~\ref{fig:b_SCS_rates_EB} (which also includes results for additional binding energies to fill in some of the gaps in $E_B$). We find that the rates are most sensitive to $E_B$ at small quarkonium three-momenta, closely following the trend of the thermal-parton number density depicted in the right panel of figure~\ref{fig:thresholds}. On the other hand, a much milder dependence is found at high momenta, as the increased available $\cm$ energy in the collision pushes down the threshold momentum substantially.  

\begin{figure}[t]
   \centering
   \includegraphics[width=\textwidth]{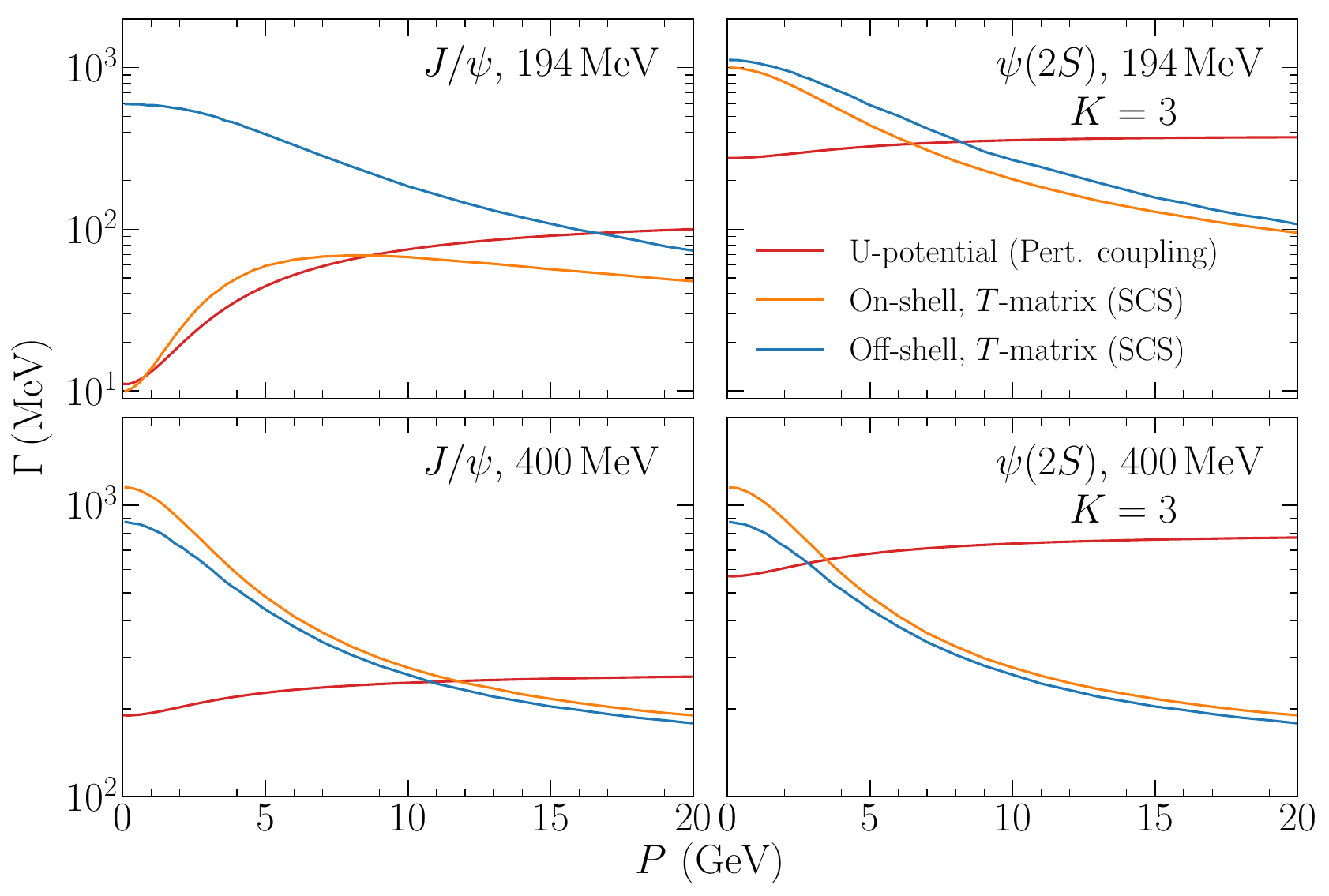}
   \caption{Dissociation rates of $J/\psi$ (left column) and $\psi(2S)$ (right column) as a function of their momenta at $T$=194 MeV (top row) and 400 MeV (bottom row). The red curves represents the on-shell $U$-potential calculation with perturbative medium coupling~\cite{Wu:2024gil} (with a $K=3$ factor for $\psi(2S)$), and the orange (blue) curves shows the SCS on-shell (off-shell) $\Tm$ results.}
    \label{fig:SCS_onshell}
  \end{figure}

In figure~\ref{fig:SCS_onshell}, we present the dissociation rates of $J/\psi$ and $\psi(2S)$ in the on-shell SCS, compared to results obtained using the $U$-potential approach with a perturbative coupling to the medium~\cite{Riek:2010fk} (see the pertinent pQCD diagrams in figure~\ref{fig:pert_diagrams}), that has been previously used in applications to quarkonium transport in URHICs at the SPS, RHIC and the LHC~\cite{Grandchamp:2001pf,Zhao:2010nk,Du:2015wha} (these calculations employ the first quasifree approach, as detailed in table~\ref{tab:qf}, with inputs shown in the left panels of figures~\ref{fig:masses} and~\ref{fig:binding}). 
We find that for $J/\psi$, the two approaches yield similar rates at $T=194\,$MeV, but deviate markedly at $T=400\,$MeV, where the binding energy of $J/\psi$ is relatively small ($\approx 50\,$MeV) and thus the effect of the nonperturbative heavy-light $T$-matrix becomes more pronounced, leading to a decreasing dependence on three-momentum---a characteristic of nonperturbative behavior.
For $\psi(2S)$ (which includes a phenomenologically determined $K$-factor of 3 in the pQCD coupling), the $U$-potential calculation yields average rates comparable to the SCS result, despite differences in momentum dependence.

A similar comparison is shown in figure~\ref{fig:bSCS_Onshell} for the rates of $\Upsilon(1S)$ and $\Upsilon(2S)$ in the SCS versus results obtained using the same inputs but with the quasifree scattering amplitudes calculated from the perturbative diagrams in figure~\ref{fig:pert_diagrams} (instead of the $T$-matrix).
The nonperturbative rates are significantly larger than the perturbative rates, particularly for states with smaller binding energies.
As temperature and momentum increase, the nonperturbative rates generally approach the perturbative results, since the $T$-matrix slowly recovers the result from one-gluon exchange, but the differences remain very large at small momenta.
Similar to  charmonia, the on-shell bottomonia rates obtained with perturbative amplitudes increase with momentum. In contrast, the on-shell rates with $\Tm$ amplitudes increase with momentum at $T=194\,$MeV, but decrease at $T=400\,$MeV.

\subsection{Off-shell effects}
\label{ssec:offshell}
%
\begin{figure}[t]
   \centering
   \includegraphics[width=\textwidth]{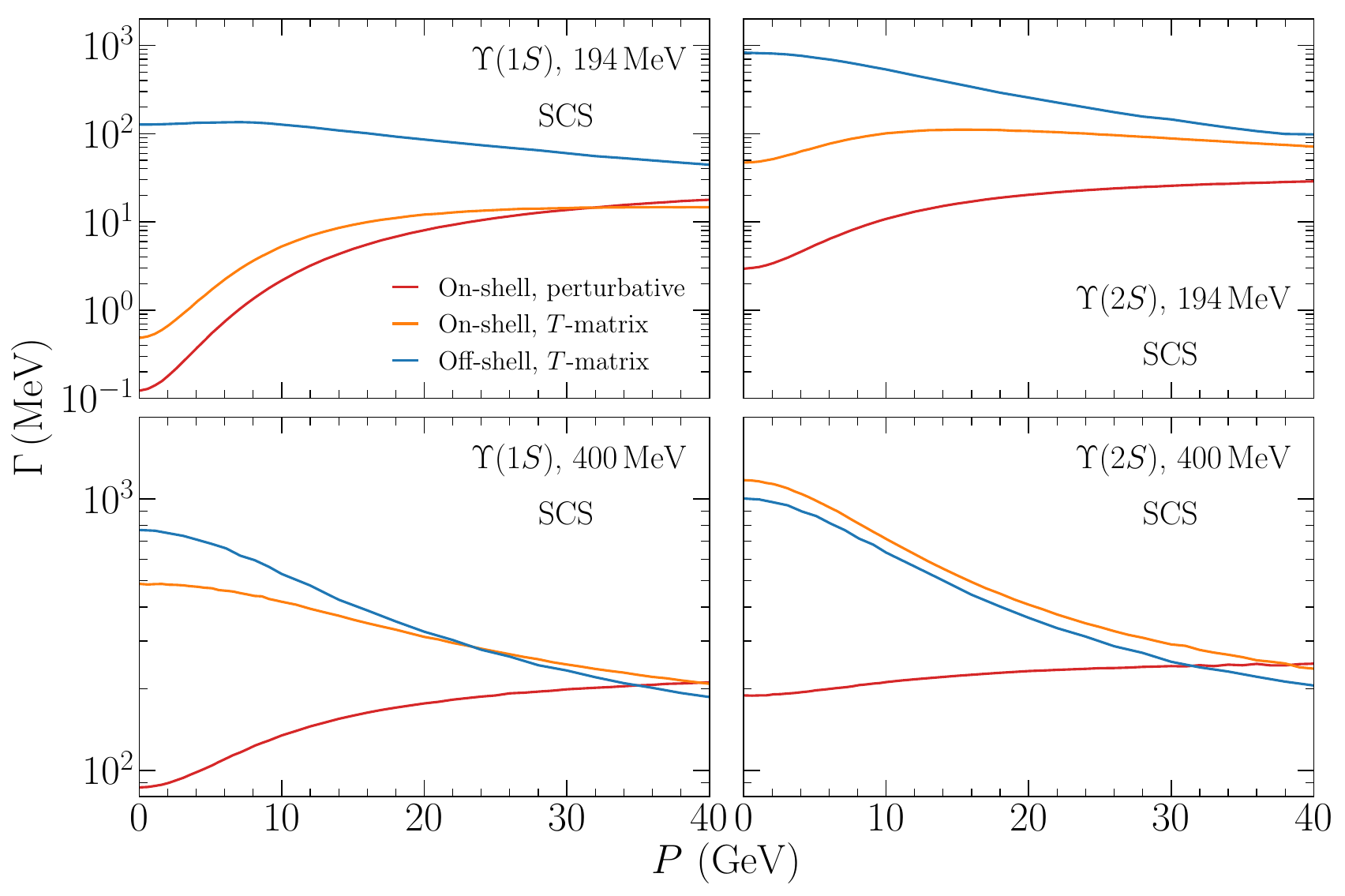}
   \caption{Dissociation rates of $\Upsilon(1S)$ (left column) and $\Upsilon(2S)$ (right column) as a function of their momenta in the SCS at $T$=194 MeV (upper) and 400 MeV (bottom). The red curves represents the on-shell perturbative calculation and the orange (blue) curves show the on-shell (off-shell) $\Tm$ results.}
    \label{fig:bSCS_Onshell}
  \end{figure}

As discussed in section~\ref{sec:hf-tmat}, the large scattering rates in an sQGP lead to broad and typically non-Lorentzian spectral functions for the partons, highlighting their off-shell properties (see figure~\ref{fig:sfs_partons}). The off-shell effects of the heavy and light partons can be implemented into the quarkonium dissociation rate via a convolution over the spectral functions of the incoming and outgoing light parton $i$ and the outgoing heavy quark $Q$ (see, \eg, \cite{Liu:2018syc}),
\begin{equation}
\begin{aligned}
   \Gamma_{\QQ}\left(\mathbf{P};T\right)= & \sum_i \frac{2}{2 \varepsilon_Q\left(\mathbf{p}_Q\right)} \int \frac{\dd \omega^{\prime} \dd^3 \mathbf{p}_Q^{\prime}}{(2 \pi)^3 2 \varepsilon_Q\left(\mathbf{p}_Q^{\prime}\right)} \frac{\dd \nu \dd^3 \mathbf{q}}{(2 \pi)^3 2 \varepsilon_i(\mathbf{q})}
   \frac{\dd \nu^{\prime} \dd^3 \mathbf{q}^{\prime}}{(2 \pi)^3 2 \varepsilon_i\left(\mathbf{q}^{\prime}\right)}\\
   &
   {(2 \pi)^4}
   \delta^{(4)}\left( p_Q+q-p_Q'-q' \right)
   \sum_{a, l, s}\overline{|\mathcal{M}_{i {Q} \rightarrow i Q}|^2}
   \rho_Q\left(\omega^{\prime}, \mathbf{p}_Q^{\prime}\right)
   \rho_i(\nu, \mathbf{q})
   \rho_i\left(\nu^{\prime}, \mathbf{q}^{\prime}\right)\\
   &\left[1-n_Q\left(\omega^{\prime}\right)\right] d_i\,n_i(\nu)
   \left[1 \pm n_i\left(\nu^{\prime}\right)\right] 
   \left[1-n_{\bar{Q}}\left(\varepsilon_{\bar{Q}}\right)\right] \,.
\end{aligned}
\label{eq:offshell}
\end{equation}
As in (\ref{eq:kinetic}), $\varepsilon_Q$,  $\varepsilon_{\Qb}$, and $\varepsilon_i$ denote the on-shell energies of the incoming heavy quark, heavy antiquark, and the light parton, respectively.
The off-shell energies of the outgoing heavy quark and the incoming and outgoing light partons are denoted by $\omega'$, $\nu$ and $\nu'$, respectively.
The spectral functions, $\rho_{Q(i)}$, describe the off-shell properties of the partons and are selfconsistently calculated within the $T$-matrix approach (recall figure~\ref{fig:sfs_partons}); they are folded over their corresponding thermal distribution functions, denoted as $n_{Q(i)}$.

The off-shell calculations of the dissociation rates for charmonia and bottomonia in the SCS are shown in figures~\ref{fig:SCS_onshell} and \ref{fig:bSCS_Onshell}, respectively.
In general, the off-shell rates are larger than the on-shell rates, particularly for the ground state at low temperatures.
This can be attributed to the reduced dissociation threshold caused by the extra strength in the parton spectral functions at energy higher than their nominal ``quasiparticle" mass. This effect relaxes kinematic constraints, effectively lowering the threshold energy squared of equation (\ref{eq:sth}).
These contributions correspond to higher-order terms in temperature, which can be physically interpreted as multiple collisions accumulating energy from the medium.
At higher quarkonium momenta and temperatures, the off-shell rates approach the on-shell results as the threshold becomes much reduced due to the increase in the (average) $\cm$ energy in the scattering of high-momentum quarkonia off thermal partons. In addition, the spectral functions at high momentum become markedly narrower~\cite{Liu:2017qah}, causing the off-shell rates to approach the on-shell rates.
For both charmonium and bottomonium, the off-shell dissociation rates at $T=400\,$MeV are lower than the on-shell rates. 
This behavior arises from a competition of two effects. On one hand, the off-shell effects lead to a higher dissociation rate because the broad spectral functions open up phase space below the quasiparticle two-body threshold, allowing for subthreshold resonance scattering~\cite{Liu:2018syc}. On the other hand, in the on-shell calculation the thermal-parton masses are reduced to make sure that the calculated QGP equation of state (EoS) still agrees with the lattice-QCD data~\cite{Liu:2018syc}. In the off-shell scenario, beyond the quasiparticle contribution, the EoS also receives contributions from parton selfenergies and two-body interactions (as characterized by the Luttinger-Ward-Baym formalism)~\cite{Liu:2017qah}, which requires larger parton masses to obtain the same pressure as from pure quasiparticles in the on-shell case. These smaller masses lead to a higher parton density, causing heavy quarkonia to interact more frequently with the medium and thus increasing the dissociation rate.


%
  \begin{figure}[t]
   \includegraphics[width=\textwidth]{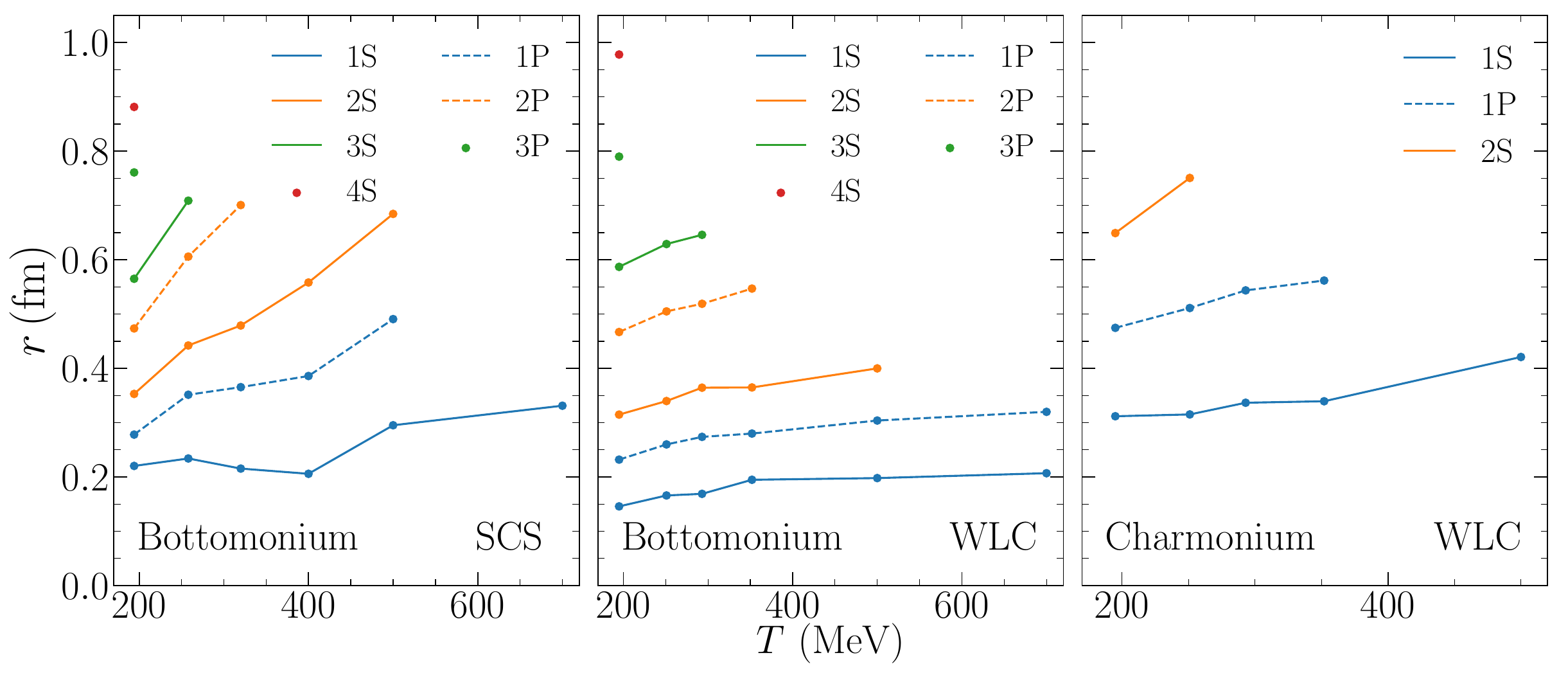}
     \caption{The in-medium radii of bottomonia in the SCS (left panel) and WLC scenario (middle panel) and of charmonia in the WLC scenario (right panel) as a function of temperature, obtained from matching the calculated dissociation rates to the (imaginary part of the) pole positions from in-medium quarkonium $T$-matrices as described in subsection~\ref{ssec:spectroscopy}. Solid and dashed lines represent the $S$-wave ($1S$, $2S$, $3S$, $4S$) and $P$-wave states ($1P$, $2P$, $3P$), respectively.}
    \label{fig:radii}
\end{figure}
%

\subsection{Interference effects}
\label{ssec:inter}
Thus far, the effect of bound quarkonium states on the quasifree dissociation has been restricted to the binding energy. However, additional wave function (or structure) effects can also affect the dissociation rates.
In particular, the size $r$ of a heavy quarkonium significantly influences its interaction with the medium. Specifically, for a colored quark
and its antiquark in close proximity, the thermal wavelength of the medium particles can become comparable to or even larger than the bound-state radius. In this regime, the colored medium partons do not effectively resolve the individual color charges of the heavy quark and antiquark.
Technically, this corresponds to an interference effect that has been elaborated in both coordinate and momentum space~\cite{Laine:2006ns,Beraudo:2007ky,Park:2007zza}, and also appears in the wave functions (or density matrices) of approaches based on open quantum systems~\cite{Brambilla:2023hkw,Delorme:2024rdo, Andronic:2024oxz}. It is sometimes also referred to as the ``imaginary part" of the $\QQb$ potential, with an $r$-dependent suppression of the incoherent rate at $r\to\infty$.
In the $\Tm$ approach, the interference effect is related to three-body effects and has been incorporated~\cite{Liu:2017qah} as an interference function, with a functional form taken from the perturbative study in \cite{Laine:2006ns}. This effect is included in the complex-pole analysis of the quarkonium properties discussed in subsection~\ref{ssec:spectroscopy}. 
In our explicit calculations of the momentum dependent rates, we also employ this functional form, introducing an ``interference factor", ($1-\exp(i\mathbf{k\cdot r})$), into equation (\ref{eq:offshell}), 
\begin{equation}
\begin{aligned}
   \Gamma_{\QQ}\left(\mathbf{P};T\right)= & \sum_i \frac{2}{2 \varepsilon_Q(\mathbf{p}_Q)} \int \frac{\dd \omega^{\prime} \dd^3 \mathbf{p}_Q^{\prime}}{(2 \pi)^3 2 \varepsilon_Q\left(\mathbf{p}_Q^{\prime}\right)} \frac{\dd \nu \dd^3 \mathbf{q}}{(2 \pi)^3 2 \varepsilon_i(\mathbf{q})}
   \frac{\dd \nu^{\prime} \dd^3 \mathbf{q}^{\prime}}{(2 \pi)^3 2 \varepsilon_i\left(\mathbf{q}^{\prime}\right)}\\
   &{(2 \pi)^4}
   \delta^{(4)}\left( p_Q+q-p_Q'-q' \right)
   \sum_{a, l, s}\overline{|\mathcal{M}_{i {Q} \rightarrow i Q}|^2}
   \rho_Q\left(\omega^{\prime}, \mathbf{p}_Q^{\prime}\right)
   \rho_i(\nu, \mathbf{q})
  \\
  & \rho_i\left(\nu^{\prime}, \mathbf{q}^{\prime}\right)\left[1-n_Q\left(\omega^{\prime}\right)\right]d_i\, n_i(\nu)
   \left[1 \pm n_i\left(\nu^{\prime}\right)\right] 
   \left[1-n_{\bar{Q}}\left(\varepsilon_{\bar{Q}}\right)\right]
   \left[ 1-e^{i \mathbf{k}\cdot \mathbf{r}} \right] \, .
\end{aligned}
\label{eq:full}
\end{equation}
where $\mathbf{k}= \mathbf{p}_Q'- \mathbf{p}_Q$ is the momentum transfer, and $\mathbf{r}$ represents the average vector separation between $Q$ and $\Qb$ (twice the average radius of the bound state) whose direction is also averaged over the azimuthal angle with respect to $\mathbf{k}$,
\begin{equation}
      \frac{1}{2}\int_0^{\pi}\left(1-e^{i k r\cos \theta}\right)\sin\theta\dd \theta = 1-\frac{\sin \left(k r\right)}{k r}\ .
         \label{eq:average_interference}
\end{equation}
In our quantitative determination of $r$ detailed in the following paragraph, the structure of the in-medium bound states is encoded in the different binding energies (and widths) of the bound states as obtained from the $T$-matrix. This includes the different sizes (or rather size distributions) of different angular-momentum states (\eg, $S$-wave vs.~$P$-wave), which are explicitly treated in the partial-wave expansion in the $T$-matrix formalism in momentum space. In particular, the low-temperature limit of the extracted radii in our calculations agrees reasonably well with the expected vacuum sizes (cf.~figure~\ref{fig:radii}).  
Note that, upon integration in equation (\ref{eq:full}), the odd powers in the argument of the exponential, $i \mathbf{k}\cdot \mathbf{r}$, vanish, ensuring that the rate remains a real quantity.
Furthermore, one also recognizes the qualitative behavior that for a large radius, the exponential oscillates rapidly, rendering a vanishing contribution; \ie, the interference effects shut off. On the other hand, for small $r$, the suppression effects become substantial, to the point that the entire interference factor, and thus the rate, vanishes as $r\to 0$.

\begin{figure}[t]
   \centering
      \includegraphics[width=\textwidth]{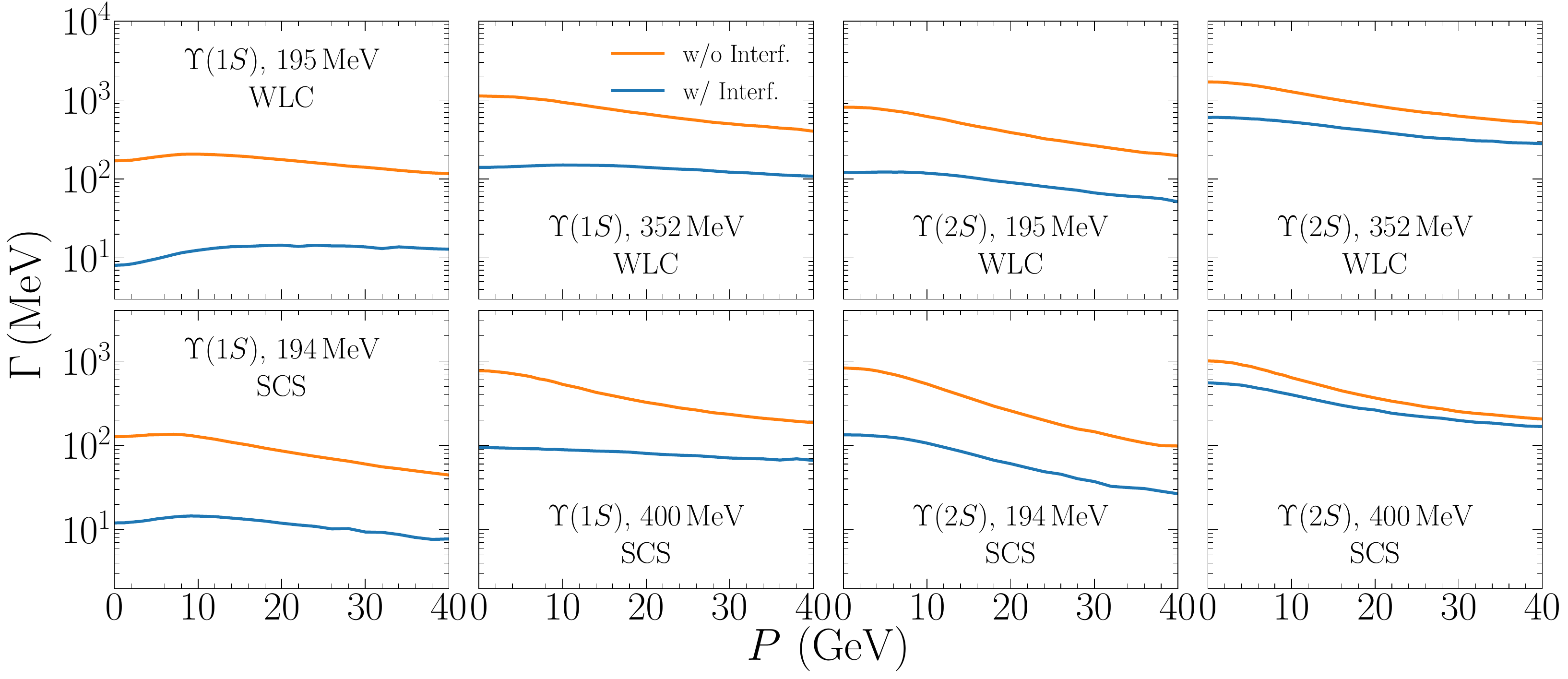}
      \caption{Illustration of the interference effect (blue lines: with, orange lines: without) in the thermal dissociation rates of bottomonia as a function of their momentum in the WLC scenario (top row) and SCS (bottom row). The left (right) two columns correspond to the $\Upsilon(1S)$  ($\Upsilon(2S)$) at different temperatures.
      }
    \label{fig:bottomonium_interference}
  \end{figure}

To quantitatively determine the average radius of each quarkonium state in a given scenario, we use the quarkonium widths and binding energies extracted from the complex-pole analysis of the quarkonium $T$-matrices as described in subsection~\ref{ssec:spectroscopy}. Concretely, we infer the in-medium radius, $r$, for each state at each temperature, so that the quarkonium dissociation rate at vanishing momentum, $\Gamma_{\QQ}(P=0)$, calculated from equation (\ref{eq:full}), matches the imaginary part of the corresponding pole of the ($S$- or $P$-wave) $\QQ$ $\Tm$ plotted in figure~\ref{fig:quarkonium_width}.
The resulting radii for bottomonia and charmonia are displayed in figure~\ref{fig:radii}.
The radii of bottomonia in the SCS (left panel) generally increase with temperature. However, for $\Upsilon(1S)$, they exhibit a non-monotonous behavior around $T\simeq400$\,MeV, which can be traced back to a similar feature in the HQ collision rates (see the left panel of figure~\ref{fig:quarkonium_width}). This behavior is ultimately related to the rather strong screening of the long-range confining force in the SCS, which reduces $b$-quark broadening but has little effect on the short-range forces that are operative within $\Upsilon(1S)$.
For the larger-size excited states, this screening leads to a noticeable increase in the size, despite the drop in collisional $c/b$-quark width.
On the other hand, in the WLC scenario, where the in-medium potential is only weakly screened, the extracted radii of both bottomonia and charmonia develop only a rather modest increase with temperature. 

%
%

In figure~\ref{fig:bottomonium_interference}, we illustrate the effect of interference on the dissociation rates of $\Upsilon(1S)$ and $\Upsilon(2S)$ at different temperatures in both SCS and WLC scenario.
As expected, the impact of interference is more pronounced for states with smaller radii but also at lower temperatures, where the thermal wavelength of the medium partons is larger, resolving less structure.
Thus, the suppression of the rate is strongest for $\Upsilon(1S)$ at the lowest temperature and quite small for $\Upsilon(2S)$ at the highest temperature, especially in the SCS, where the screening increases markedly with $T$. In addition, higher momenta of the quarkonia, leading on average to larger momentum transfers in the collisions with thermal partons, enable a better resolution of their structure and thus the interference becomes weaker.

\begin{figure}[t]
   \centering
     \includegraphics[width=\textwidth]{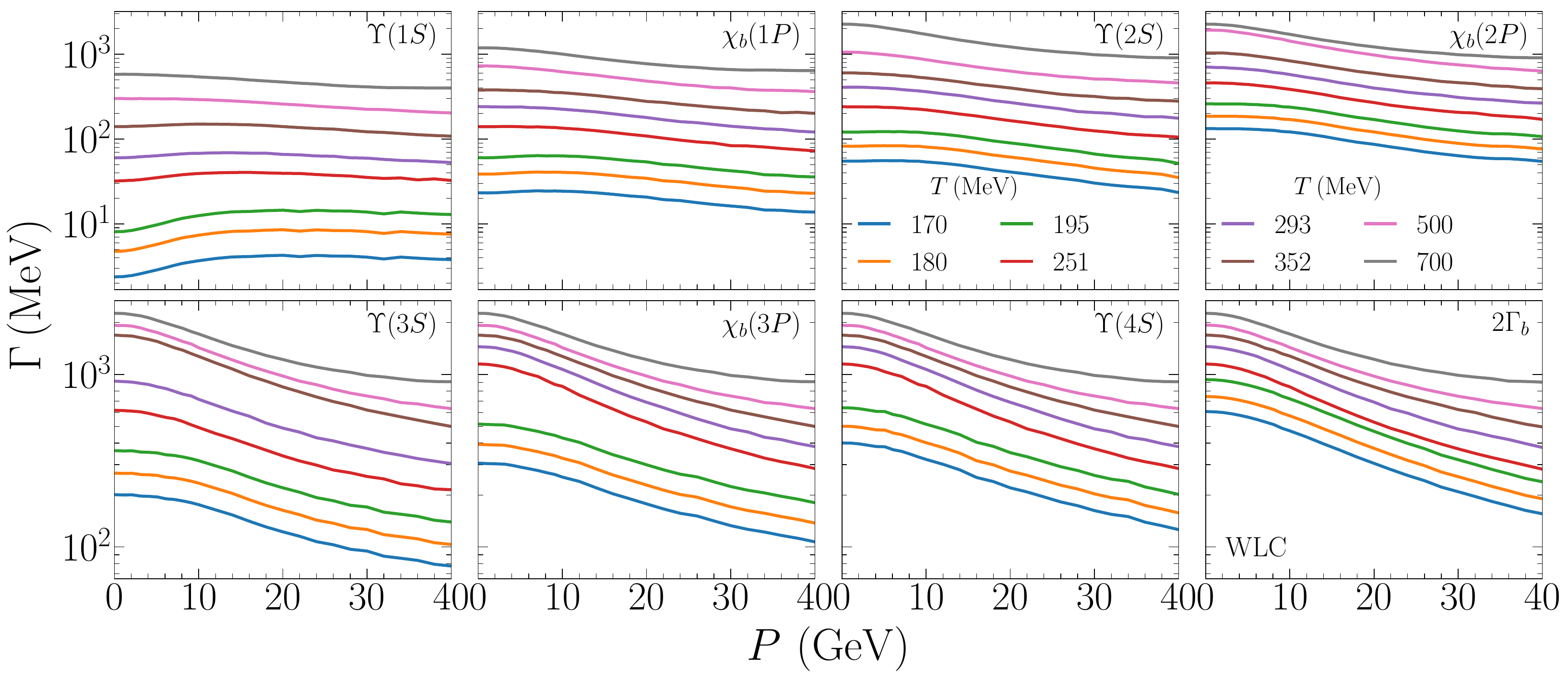}
     \caption{Compilation of our final results for the thermal dissociation widths of all bottomonium states in the WLC scenario considered in this work , as a function of their momenta for the same 8 different temperatures (different colors) in each panel. From left to right, and top to bottom, the panels are organized by the decreasing binding energy of the states:  $\Upsilon(1S)$,  $\chi_b(1P)$, $\Upsilon(2S)$, $\chi_b(2P)$, $\Upsilon(3S)$, $\chi_b(3P)$, and $\Upsilon(4S)$.  The last panel (lower right) shows two times the collisional width of the bottom-quark, $\Gamma_b$, corresponding to the limit of vanishing binding and infinite radius.}
    \label{fig:brates_p}
  \end{figure}

In figure~\ref{fig:brates_p}, we collect the final results of thermal rates of all bottomonium states considered in this work (off-shell with interference and binding-energy effects), as a function of momentum at different temperatures within the WLC scenario.
In addition to the expected ordering by binding energy and radius (where both large binding and small radius reduce the rates), one notices subtle changes in the three-momentum dependence, which exhibits an increasing trend for compact strongly bound states (due to larger interference and reduced phase space at small momenta), gradually transitioning into a rather strongly decreasing trend.
The latter is inherent in the limiting case of twice the collisional width of $b$-quarks, which, in turn, emerges from predominant nonperturbative effects at low $P$ caused by the long-range confining force, transitioning to perturbative behavior governed by a rather weak but short-range color-Coulomb interaction at high $P$.

\begin{figure}[t]
     \includegraphics[width=\textwidth]{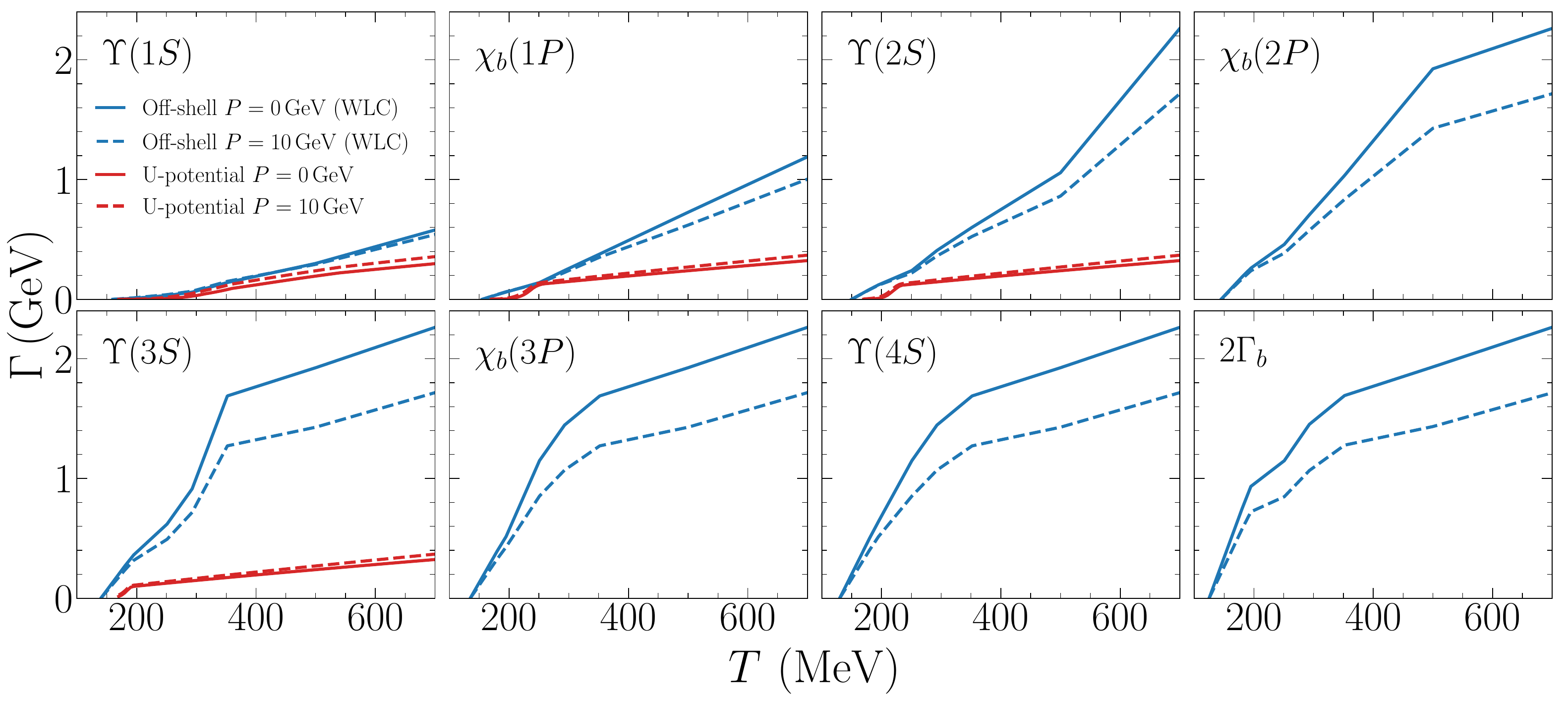}
     \caption{Our final results for the bottomonium rates as a function of temperature in the WLC scenarios (blue lines, including $1S$, $2S$, $3S$, $4S$, $1P$, $2P$, and $3P$ ), compared to the rate calculations based on  the $U$-potential (on-shell and with perturbative amplitudes) in \cite{Du:2017qkv} (red lines, including $1S$, $2S$, $3S$, and $1P$). The last panel shows twice the bottom-quark collisional rate in the WLC scenario (it is much smaller with perturbative amplitudes). Solid (dashed) lines represent the calculations for $P = 0(10)$\,GeV.}
    \label{fig:brates_T}
\end{figure}

In an alternative projection, we display in figure~\ref{fig:brates_T} the bottomonium rates from figure~\ref{fig:brates_p}, as a function of temperature at two fixed momenta ($0$ and $10\,$GeV), comparing the WLC scenario with perturbative rates from previous transport calculations~\cite{Du:2017qkv} (for the available states).
For the latter rates, their inputs are shown in the left panels of figures~\ref{fig:masses} and \ref{fig:binding}, and are coupled to a quasiparticle medium via the pQCD amplitudes shown in figure~\ref{fig:pert_diagrams}.
For the $\Upsilon(1S)$, the results are comparable, especially at temperatures below $\sim$400\,MeV, mostly thanks to the strong interference effect. 
Also, for the first-excited states, $\Upsilon(2S)$ and $\chi_b(1P)$, one finds approximate agreement at low $T$, but the nonperturbative amplitudes and large collisional parton widths cause strong deviations from the previous results at higher temperatures, $T\gsim300$\,MeV. For the $\Upsilon(3S)$ this trend is further reinforced.
The rates at finite momentum are indicated in the difference
between the solid and dashed lines. The nonperturbative rates at higher momenta are generally smaller, whereas the perturbative rates exhibit the opposite trend, although the differences are not large, in part because a momentum of 10\,GeV is still relatively small, being comparable to the rest masses.
%

\subsection{Comparison of approximation schemes}
\label{ssec:approx}

A low-energy effective field theory (EFT) of quarkonia at finite temperature, applicable in the regime where $m_Q \gg \frac{1}{r}\gg E_B$, with $r$ being the quarkonium size, can be formulated based on potential non-relativistic QCD (pNRQCD)~\cite{Brambilla:1999xf}.
The hierarchy $m_Q \gg \frac{1}{r}$ ensures that the heavy quark is non-relativistic, while $\frac{1}{r}\gg E_B$ allows the interaction between the heavy quark and the antiquark to be described by a potential at leading order in $r E_B$.
Applying this to finite temperature in the regime where $1/r\gg \pi T, m_D \gg E_B$~\cite{Brambilla:2016wgg}, the quarkonium dissociation rate can be expressed as the product of the momentum diffusion coefficient $\kappa$ and the squared quarkonium size, 
\begin{equation}
   \Gamma_{\QQ} = \kappa r^2 \ .
   \label{eq:rate_EFT}
\end{equation}
To connect this result to our calculation, we expand the interference term in equation (\ref{eq:full}) as
\begin{equation}
      1-\exp\left(i \mathbf{k}\cdot \mathbf{r}\right)=1-\cos\left(\mathbf{k}\cdot \mathbf{r}\right)-i\sin\left(\mathbf{k}\cdot \mathbf{r}\right) \ .
        \label{eq:interference_function}
\end{equation}
The imaginary part of equation (\ref{eq:interference_function}) vanishes upon integration in (\ref{eq:full}), since the momentum transfer $\mathbf{k}$ is symmetric with respect to the orientation of $\mathbf{r}$.
Expanding the real part, we obtain
\begin{equation}
    \rm{Re}\left[1-\exp\left(i \mathbf{k}\cdot \mathbf{r}\right)\right]=\frac{1}{2}\left(\mathbf{k}\cdot \mathbf{r}\right)^2
-\frac{\left(\mathbf{k}\cdot \mathbf{r}\right)^4}{24}+\mathcal{O}\left(\left(\mathbf{k}\cdot \mathbf{r}\right)^6\right) \ .
    \label{eq:int_expansion}
\end{equation}
We align the direction of the dipole vector $\mathbf{r}$ with the momentum of the incoming heavy quark so that,
to nontrivial leading order, equation (\ref{eq:int_expansion}) becomes
\begin{equation}
       \frac{1}{2}\left(\mathbf{k}\cdot \mathbf{r}\right)^2=\frac{1}{2}\mathbf{p}_Q^2 r^2 \left(1-\frac{\mathbf{p}_Q'\cdot \mathbf{p}_Q }{\mathbf{p}_Q^2}\right)^2
      = r^2 \frac{1}{2}\left[\mathbf{p}_Q^2-2\mathbf{p}_Q'\cdot \mathbf{p}_Q +\frac{\left(\mathbf{p}_Q'\cdot \mathbf{p}_Q\right)^2}{\mathbf{p}_Q^2}\right]\, .
   \label{eq:kr2}
\end{equation}
In doing so, the LO expansion of the rate \eqref{eq:full} becomes proportional to the {\em longitudinal} HQ momentum diffusion coefficient~\cite{Rapp:2009my},
\begin{equation}
B_1(\mathbf{p}_Q) =\frac{1}{2}\left[\mathbf{p}_Q^2\langle 1\rangle-2\left\langle\mathbf{p}_Q^{\prime} \cdot \mathbf{p}_Q\right\rangle+\frac{\left\langle\left(\mathbf{p}_Q \cdot \mathbf{p}_Q^{\prime}\right)^2\right\rangle}{\mathbf{p}_Q^2}\right] \ ,
\label{eq:kappa}
\end{equation}
where 
\begin{equation}
\begin{aligned}
   \langle X\rangle \equiv & \sum_i \frac{2}{2 \varepsilon_Q(\mathbf{p}_Q)} \int \frac{\dd \omega^{\prime} \dd^3 \mathbf{p}_Q^{\prime}}{(2 \pi)^3 2 \varepsilon_Q\left(\mathbf{p}_Q^{\prime}\right)} \frac{\dd \nu \dd^3 \mathbf{q}}{(2 \pi)^3 2 \varepsilon_i(\mathbf{q})}
   \frac{\dd \nu^{\prime} \dd^3 \mathbf{q}^{\prime}}{(2 \pi)^3 2 \varepsilon_i\left(\mathbf{q}^{\prime}\right)}\\
   &{(2 \pi)^4}
   \delta^{(4)}\left( p_Q+q-p_Q'-q' \right)
   \sum_{a, l, s}\overline{|\mathcal{M}_{i {Q} \rightarrow i Q}|^2}
   \rho_Q\left(\omega^{\prime}, \mathbf{p}_Q^{\prime}\right)
   \rho_i(\nu, \mathbf{q})
   \rho_i\left(\nu^{\prime}, \mathbf{q}^{\prime}\right)\\
   &\left[1-n_Q\left(\omega^{\prime}\right)\right] d_i n_i(\nu)
   \left[1 \pm n_i\left(\nu^{\prime}\right)\right] 
   \left[1-n_{\bar{Q}}\left(\varepsilon_{\bar{Q}}\right)\right] X \ ,
\end{aligned}
\label{eq:thermal_average}
\end{equation}
so that (\ref{eq:full}) becomes
\begin{equation}
      \Gamma_{\QQ}\left(\mathbf{P} ; T\right)=\kappa\left(\frac{\mathbf{P}}{2}\right)r^2 \ ,
      \label{eq:eft}
\end{equation}
with $\kappa\left(\frac{\mathbf{P}}{2}\right)$ evaluated at half of the quarkonium momentum.
Note, however, that in our original expression, the kinematics of the heavy-light scattering amplitude includes the quarkonium binding energy, while the latter is not present for the (on-shell) HQ diffusion coefficient.

\begin{figure}[t]
   \centering
\includegraphics[width=\textwidth]{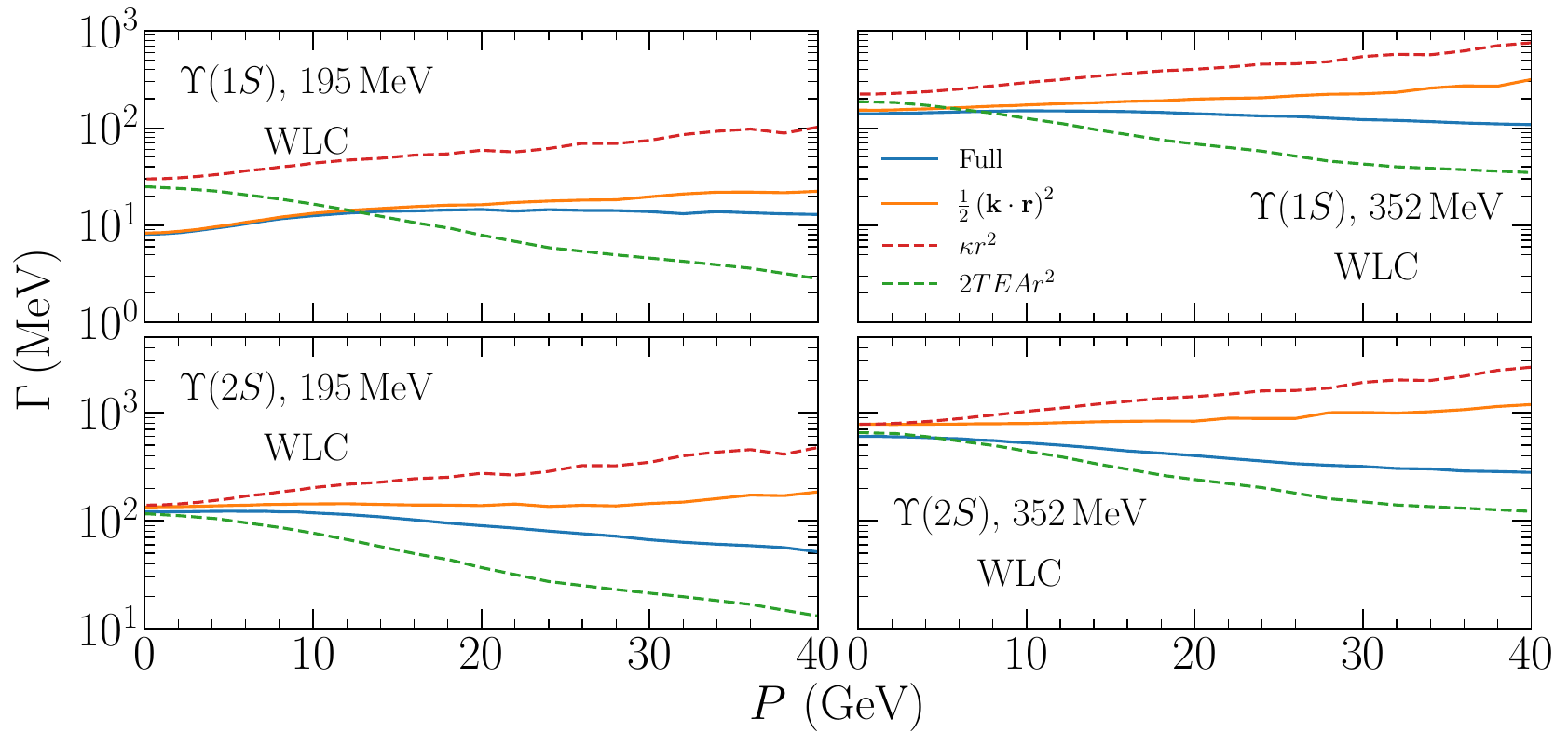}
\caption{Thermal bottomonium widths as a function of momentum in the WLC scenario. The top (bottom) row corresponds to the $\Upsilon(1S)$ ($\Upsilon(2S)$) at temperatures of $195\,$MeV (left column) and $352\,$MeV (right column). The blue lines represent our full calculations, the orange lines show the result of the expansion of the interference effect to leading order in $(\mathbf{k} \cdot \mathbf{r})^2$, the red dashed lines correspond to $\kappa r^2$, where $\kappa$ is the $b$-quark momentum diffusion coefficient, and the green dashed lines result from expressing $\kappa$ in terms of the HQ friction coefficient, $A$.}
    \label{fig:b_rates_expansion}
\end{figure}

In figure~\ref{fig:b_rates_expansion} we compare the momentum dependence of
bottomonium dissociation rates at two temperatures, $T = 195, 352\,\text{MeV}$, for our full calculation (WLC scenario with interference effect) to its LO expansion and the EFT approximation, equation (\ref{eq:eft}).
At small momentum and low temperatures (left panels), where the radii are relatively small, the LO expansion closely follows the full result for both $1S$ and $2S$ states.
However, as the momentum transfer, $\mathbf{k}$, increases, the LO expansion begins to deviate from the full calculation, since neglecting the next-to-leading order correction, $-(\mathbf{k}\cdot \mathbf{r})^4/{24}$,
causes the LO approximation to overestimate the full calculation. This becomes more pronounced at higher temperatures.
The EFT approximation was, in principle, only developed for small momenta, but it neglects the effect of binding energy and therefore overestimates the full rates for the $\Upsilon(1S)$ at low $T$.
At higher $T$, and for the $\Upsilon(2S)$ at low $T$, the agreement is fair. These observations are consistent with the criterion $1 \gg r E_B$, which holds when both the quarkonium binding energy and radius are relatively small.
When inspecting the three-momentum dependence of the EFT result, one should recall a well-known issue with the longitudinal HQ momentum diffusion coefficient: it can lead to rather large violations of the fluctuation-dissipation theorem (FDT)~\cite{Rapp:2009my}.
In the HQ diffusion context, the FDT is usually enforced to leading order in 1/$m_Q$ by adjusting $B_1$ to the friction coefficient, $A$, via $B_1=TE_Q A$ (which is corroborated by studies with the Boltzmann equation~\cite{Rapp:2018qla}).
We therefore also show results for the EFT approximation when expressing $\kappa$ through $A$. This essentially changes the three-momentum dependence from increasing to decreasing and aligns better with the full calculations for small binding energies.  

%
\begin{figure}[t]
\includegraphics[width=\textwidth]{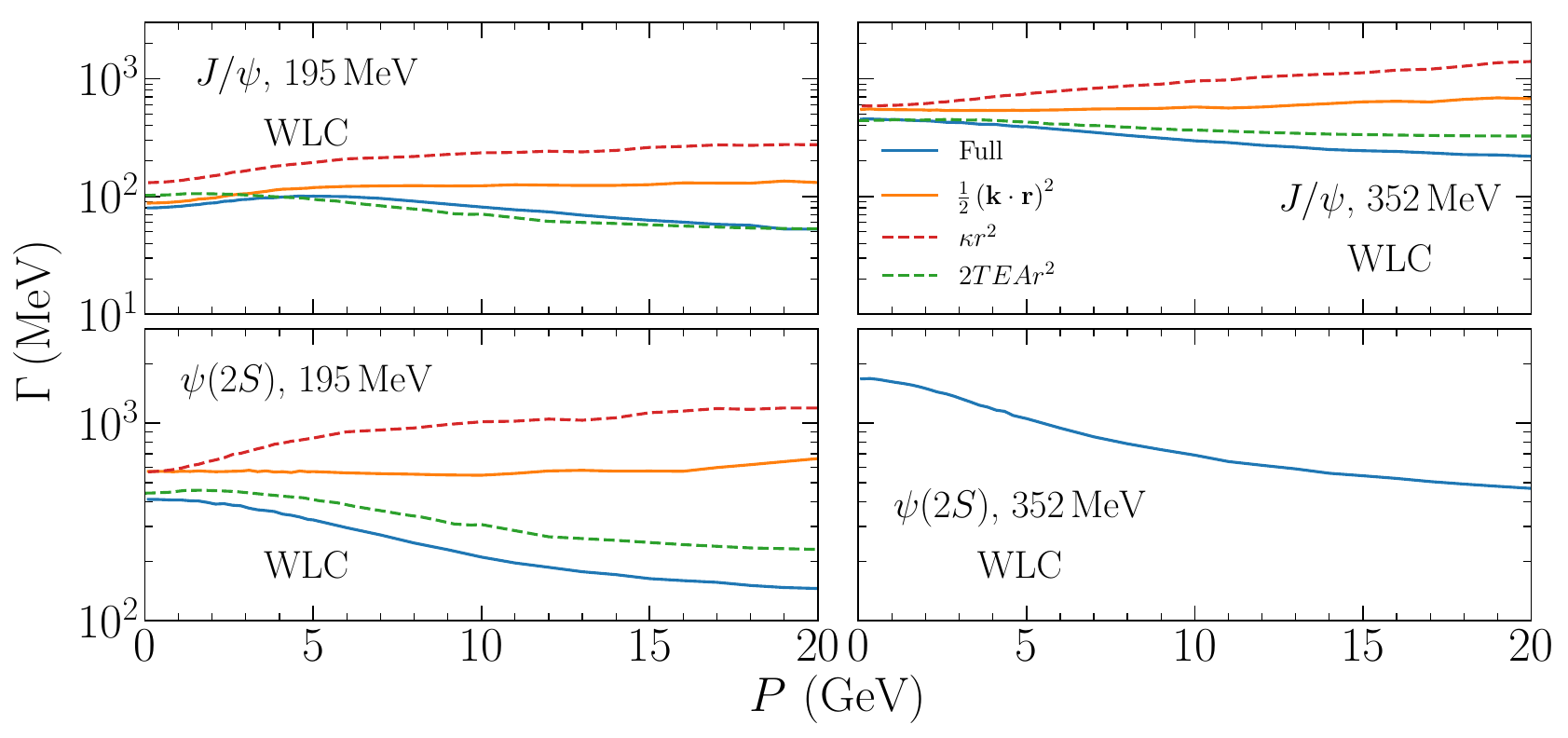}
\caption{Same as figure~\ref{fig:b_rates_expansion} but for charmonia, \ie, $J/\psi$ (upper row) and $\psi(2S)$ (lower row).
}
    \label{fig:c_rates_expansion}
\end{figure}

Similar behaviors are observed for charmonia, see figure~\ref{fig:c_rates_expansion}. Here, the expansions are closest to the full calculation for the $J/\psi$ (which has binding energies comparable to the $\Upsilon(2S)$) while for the $\psi(2S)$ the relatively large radius compromises the small-radius assumption.

\section{Conclusions}
\label{sec:concl}
We have investigated the dissociation rates of quarkonia in the quark-gluon plasma, with emphasis on elaborating the nonperturbative effects that are believed to underlie the strong-coupling properties of the partonic medium. We have chosen the thermodynamic $\Tm$ approach as our theoretical framework, previously developed to achieve a consistent description of the spectral and transport properties of the QGP, embedded in a realistic equation of state.
Focusing on inelastic scattering processes as the main contributor to quarkonium dissociation, and starting from an earlier employed quasifree approximation, we have systematically assessed the effects of in-medium quarkonium binding energies, nonperturbative heavy-light scattering amplitudes, broad partonic spectral function (as opposed to on-shell quasiparticle approximations) and quarkonium structure, on the dissociation rates. In particular, we have taken advantage of a recent pole analysis to extract effective binding energies and radii to ensure a selfconsistent treatment of quarkonium, heavy-quark and thermal-parton properties.
Specifically, the nonperturbative heavy-light interactions, together with the broad spectral functions which mitigate the threshold effects from (large) binding energies, lead to a substantial increase of both charmonium and bottomonium reaction rates compared to previous calculations with a perturbative coupling to a quasiparticle medium, especially for excited states. Quarkonium structure effects, which cause a destructive interference in the inelastic scattering off heavy quarks and antiquarks, lead to a large suppression of the rates for the ground states ($J/\psi$ and $\Upsilon(1S)$), but also for the excited states at low temperatures. Another noteworthy feature is the three-momentum dependence of the rates, which exhibits an interesting transition from increasing with momentum for strongly bound states to a decreasing momentum dependence for weakly bound states, reflecting the nonperturbative behavior of the collision rates of the individual heavy quarks.
Our set-up also allowed us to test the often used dipole approximation; we found it to be fairly reliable if the binding energy of the state is small, and at small momenta. 

Moving forward, we plan to implement the new rates, which are now firmly rooted in the physics of the sQGP and quantitatively constrained by a variety of lQCD data,
into phenomenological applications to heavy-ion collisions. We anticipate that for the ground states, previous findings might uphold, but the very large rates, especially for the excited bottomonia, may lead to a significant change in the composition of their production yields: the larger rates will likely lead to a larger suppression of primordially produced quarkonia, while simultaneously enhancing the regeneration reactions. Work in this direction is underway. 

\section*{Acknowledgments}
This work is supported by the U.S. National Science Foundation under grant no. PHY-2209335 and the Topical Collaboration in Nuclear Theory on \textit{Heavy-Flavor Theory (HEFTY) for QCD Matter} under award no.~DE-SC0023547.

\bibliographystyle{JHEP}
\bibliography{refcnew}

\end{document}